\documentclass[%
 aip,%
 pop,%
 amsmath,%
 amssymb,%
 preprint,%
]{revtex4-1}

\usepackage{dcolumn}
\usepackage{bm}
\usepackage[super]{nth}
\usepackage{graphicx}
\usepackage{caption,subcaption}
\usepackage[%
pdfauthor={Peter Hill},%
pdftitle={FCI in BOUT++},%
pdftex]{hyperref}
\usepackage{cleveref}

\newcommand{\norm}[1]{\|#1\|}
\newcommand{\myvec}[1]{\vec{#1}}
\newcommand{\bout}[0]{\textit{\textsc{BOUT++}}}
\newcommand{\zoidberg}[0]{\textit{\textsc{Zoidberg}}}
\DeclareMathOperator{\gauss}{gauss}

\begin{document}

\preprint{AIP/123-QED}

\title{Dirichlet boundary conditions for arbitrary-shaped boundaries in stellarator-like magnetic fields for the Flux-Coordinate Independent method}
\author{Peter Hill}
\email{Peter.Hill@york.ac.uk}
\author{Brendan Shanahan}
\author{Ben Dudson}
\affiliation{York Plasma Institute, University of York}

\date{\today}

\begin{abstract}
  We present a technique for handling Dirichlet boundary conditions with the Flux Coordinate Independent (FCI) parallel derivative operator with arbitrary-shaped material geometry in general 3D magnetic fields.
The FCI method constructs a finite difference scheme for $\nabla_\Vert$ by following the field lines between poloidal planes and interpolating within planes, rather than having a field-aligned mesh on flux surfaces.
Doing so removes the need for field-aligned coordinate systems that suffer from singularities in the metric tensor at null points in the magnetic field (or equivalently, when $q \to \infty$).
One cost of this method is that as the field lines are not on the mesh, they may leave the domain at any point between neighbouring planes, complicating the application of boundary conditions.

The Leg Value Fill (LVF) boundary condition scheme presented here involves an extrapolation/interpolation of the boundary value onto the field line end point.
The usual finite difference scheme can then be used unmodified.
We implement the LVF scheme in \bout{} and use the Method of Manufactured Solutions to verify the implementation in a rectangular domain, and show that it doesn't modify the error scaling of the finite difference scheme.
We outline the use of LVF for arbitrary wall geometry.

We also demonstrate the feasibility of using the FCI approach in non-axisymmetric configurations for a simple diffusion model in a ``straight stellarator'' magnetic field.
A Gaussian blob diffuses along the field lines, tracing out flux surfaces.
Dirichlet boundary conditions impose a last closed flux surface (LCFS) that confines the density.
Including a poloidal limiter moves the LCFS to a smaller radius.

The expected scaling of the numerical perpendicular diffusion, which is a consequence of the FCI method, in stellarator-like geometry is recovered.
A novel technique for increasing the parallel resolution during post-processing, in order to reduce artefacts in visualisations, is described.
\end{abstract}

\maketitle

\section{Introduction}
\label{sec:intro}

Anisotropic phenomena are prevalent in magnetised plasmas.
The Lorentz force tends to confine charged particles to magnetic field lines, with the result that the characteristic size of spatial variations of macroscopic plasma quantities are larger in the direction parallel to the magnetic field compared to those in the perpendicular plane.

Computational techniques take advantage of this anisotropy by, for example, aligning the computational grid to the magnetic field and reducing the resolution in the parallel direction.
However, field-aligned coordinate systems typically have difficulties handling changes in magnetic topology; X-points, for instance, introduce singularities in the metric tensor.
The Flux Coordinate Independent (FCI) parallel derivative operator\cite{Hariri2013,Hariri2014,Hill2015,Stegmeir2016} does not require a field-aligned coordinate system, allowing the use of simpler grids in the perpendicular plane while still allowing efficient handling of anisotropic physics.

In this work, we extend the FCI technique to handle arbitrarily shaped boundaries, including limiters, and demonstrate its use in stellarator-like fields.
This work is organised as follows:
in \cref{sec:fci}, we explain the FCI method and discuss its implementation;
in \cref{sec:interp,sec:3d-fields}, we discuss some issues about interpolation and non-axisymmetric magnetic fields;
simulations of stellarator-like magnetic fields are in \cref{sec:sims}.
We also describe a novel technique for upscaling visualisations in \cref{sec:upscale}.

\section{Flux-Coordinate Independent method for parallel derivatives}
\label{sec:fci}

Conventionally in magnetised plasma turbulence simulations, derivatives parallel to the magnetic field are taken by using a field-aligned coordinate system.
However, these are tied to flux surfaces, and hence suffer from inevitable singularities in the metric tensor when attempting to encompass multiple magnetic topologies, i.e. crossing separatrices.
These singularities can be numerically challenging to handle.

\begin{figure}[hb]
  \centering
  \includegraphics[width=0.5\textwidth]{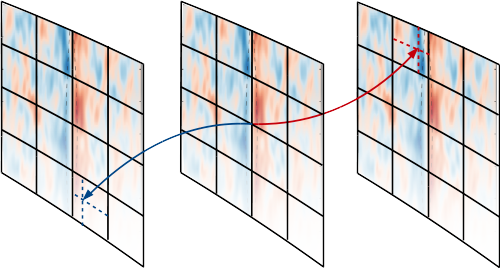}
  \caption[FCI schematic]{Schematic of the Flux Coordinate Independent method for the parallel derivative operator.
Starting from a given grid point, magnetic field lines are traced in the forward and backward directions.
The argument of the operator is interpolated to find the value at the location where the field line intersects the adjacent perpendicular slices, allowing a finite difference scheme to be constructed.}
  \label{fig:fci-schematic}
\end{figure}

The Flux-Coordinate Independent (FCI) method for the parallel derivatives of a function is conceptually simple:
one first follows the magnetic field line from a given grid point in both directions until it intersects the two adjacent perpendicular planes (see \cref{fig:fci-schematic}).
The function to be differentiated is then interpolated in the perpendicular plane at the field intersection points, and a finite difference scheme can be constructed using these values and the value at the emitting grid point.
Higher order finite difference schemes may be constructed by following the field line past further perpendicular planes, interpolating at each intersection point.
It should be noted at this point that while FCI is strictly formulated on perpendicular planes, in practice, poloidal planes are often used.
This is a reasonable approximation, given the assumptions of strong anisotropy required by FCI, and we use the terms ``perpendicular'' and ``poloidal'' interchangeably throughout this work.

As the finite difference scheme is constructed at each individual grid point, the coordinate system in the perpendicular plane is no longer tied to the flux surfaces and in principle any mesh may be used.
Other concerns may limit the choice of mesh, e.g. the need for easy flux-surface averages, which may require a flux-surface mesh in part of the plasma.
Another consideration is that while it is possible to vastly drop the resolution in the parallel direction (i.e. the inter-plane spacing) with only a small loss in accuracy, similar to conventional field-aligned grids, one must still retain enough resolution in the perpendicular mesh to capture the relevant physics of interest.

\subsection{Comparison with the standard \bout{} mesh}
\label{sec:boutmesh-comparison}

\bout{}\cite{Dudson2009,Dudson2014b,Dudson2016} is a free and open source framework designed to solve partial differential equations, with an emphasis on models of magnetically confined plasmas.
It has been used for a variety of applications, from edge\cite{Xi2012,Dudson2011,Snyder2011} and scrape-off layer\cite{Walkden2013,Angus2012} physics in tokamaks, to turbulence in linear devices\cite{Friedman2012,Shanahan2014}.

\bout{} discretises space on a three-dimensional mesh, with the dimensions labelled $x$, $y$ and $z$.
Typically, $x$ is the ``radial'' direction, $y$ the ``poloidal'', and $z$ the ``toroidal''.
The conventional ``ballooning''-style \bout{} coordinate system\cite{Xu2008,Dudson2009}, for $\psi, \theta, \zeta$ the usual orthogonal tokamak coordinates, is defined as:
\begin{align}
  \label{eq:bout-coords}
  x = \psi, \qquad
  y = \theta, \qquad
  z = \zeta - \int_{\theta_0}^{\theta}\nu d\theta,
\end{align}
where $\nu$ is the local field line pitch, given by
\begin{equation}
  \label{eq:pitch-angle}
  \nu(\psi, \theta) = \frac{\partial \zeta}{\partial \theta} = \frac{\myvec{B}\cdot\nabla\zeta}{\myvec{B}\cdot\nabla\theta}.
\end{equation}

By keeping $z$ fixed and moving in $y$, the integral in $z$ changes so we need to move in $\zeta$.
This moves us along a field line.
Essentially, $y$ is the coordinate along the field line while $z$ picks out different field lines.
Because the physics of interest are expected to be field-aligned, we are able to use a lower resolution in $y$ and still resolve the physical scales.

The metric tensor for this coordinate system is orthogonal only at one $y$-location, meaning as we move in $y$, cross-terms appear in the $x$-derivatives.
It is possible to eliminate these cross-terms by applying a shifted metric\cite{Scott2001,Hariri2013}.
To do this, at each $y$-point, we can shift $z$ by the integral in \cref{eq:bout-coords}, effectively moving us back into non-field-aligned coordinates, performing the derivatives in $x$, and then transforming back to the field-aligned coordinates.
This can be done using Fast Fourier Transforms (FFTs) which are computationally inexpensive.

At either $y$-end of the grid we need to shift in $z$ in order to match the field lines in a twist-shift boundary\cite{Dimits1993}.
This needs to be done regardless of whether or not we choose to use the shifted metric to eliminate the $x$-derivative cross-terms.

In contrast to the standard \bout{} coordinate system, the FCI method explicitly does not use field-aligned coordinates.
The construction of the parallel derivatives in fact has the major advantages of a field-aligned system (reduced resolution in the parallel direction) but allows more freedom in the choice of coordinates for the perpendicular directions.
For example, two possible choices of coordinate system are tokamak coordinates:
\begin{align}
  \label{eq:fci-coords}
  x &= \psi, \qquad
  y = \zeta, \qquad
  z = \theta,
\end{align}
or cylindrical coordinates:
\begin{align}
  x &= R, \qquad
  y = \zeta, \qquad
  z = Z.
\end{align}

Internally, \bout{} assumes that $y$ is the ``parallel'' direction, for e.g. communication.
FCI requires $\zeta$ to be the ``parallel'' direction, thus the FCI implementation in \bout{} identifies $y \equiv \zeta$.
Another way of looking at this is that the usual \bout{} mesh identifies $y$ with the poloidal direction whereas the FCI mesh identifies it with the axisymmetric (or guide field) direction.

FCI inherently employs a shifted metric, so no cross-terms appear in the the perpendicular derivatives, simplifying the calculations, and no twist-shift has to be performed.

While it is technically possible to switch between using the standard \bout{} mesh and FCI for a given problem, currently there are some technical hurdles.
The assumptions on the nature of $x$, $y$, and $z$ in \bout{} simultaneously limit FCI in the choice of perpendicular coordinates, while lifting some restrictions in the parallel direction.
The current implementation of \bout{} assumes that $z$ is axisymmetric, but makes no such assumption on $y$.
Thus, using FCI, it is possible to simulate non-axisymmetric configurations, such as stellarators, which are not possible otherwise, at the cost of complicating the inclusion of curvature effects.
Note that these obstacles are not inherent to FCI -- merely the implementation of FCI in \bout{}.
Overcoming these technical limitations is the focus of future work.

\subsection{Boundary conditions}
\label{sec:boundaries}

\subsubsection{Simple geometry}
\label{sec:simple-geom}

While FCI has already been implemented in other codes\cite{Hariri2013,Hariri2014,Stegmeir2016} and used for plasma simulations\cite{Hill2015}, the boundaries of the simulation domain were either periodic, or treated very simply.
The problem is how to treat field lines correctly when they intersect with or leave the simulation boundaries.
For example, in Ref.~\onlinecite{Hariri2014}, the magnetic topology was a cylinder, and a mask was applied to the simulation domain such that the equations were not solved outside of a radius $r$.
A different solution was used in Ref.~\onlinecite{Hill2015}, where the simulation was periodic in two directions, and the component of the magnetic field in the third direction was damped close to the edges, such that the resulting field was tangential to the edge.
Field lines then never intersected the domain boundaries, and boundary conditions could be applied in the perpendicular direction only.

\begin{figure}[ht]
  \centering
  \includegraphics[width=0.4\textwidth]{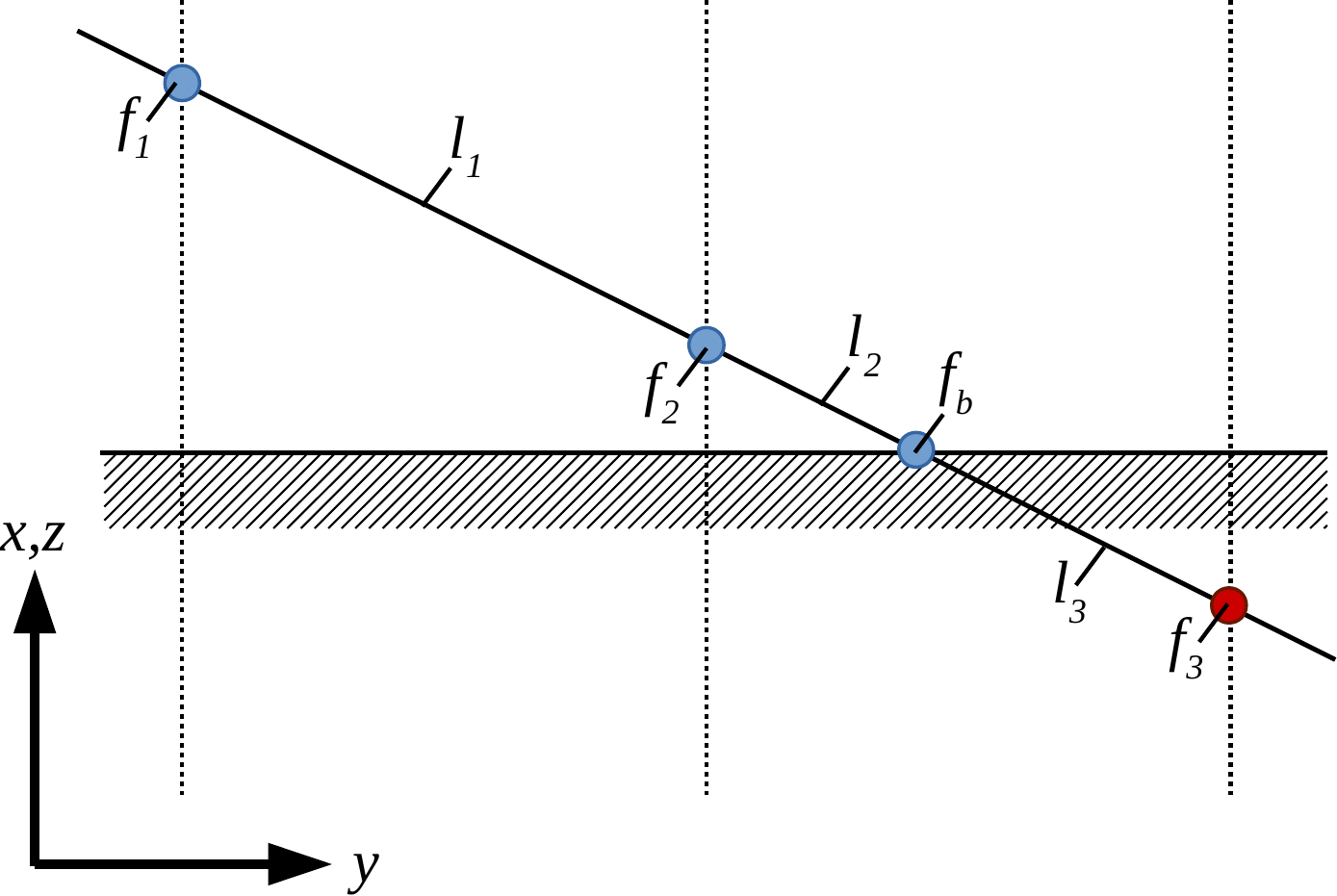}
  \caption{A field line leaving the boundary.
$f_2$ is located on a grid point, while $f_1,f_3$ are located on intersection points with the adjacent perpendicular planes, and $f_b$ is located on the intersection with the boundary.
$l_1, l_2, l_3$ are distances along the field lines between the four points above.}
  \label{fig:problem-desc}
\end{figure}

Let us first consider a scalar field $f$ on a simple, uniform, rectangular grid with boundaries located at half the grid spacing outside the first and last points in each of the grid dimensions.
For any given point in the grid where the field line traced from this point intersects the boundary before intersecting the next perpendicular plane, we need to be able to calculate parallel derivatives.
This situation is depicted in \cref{fig:problem-desc}, where $f_2$ is the value of the scalar field at the point in question, $f_1$ and $f_3$ are the values at the intersection points with the adjacent perpendicular planes in the negative and positive $y$ directions, respectively; $f_b$ is the value on the boundary; $l_{1,2,3}$ are the parallel distances between $f_{1,2}, f_{2,b}, f_{b,3}$ respectively.

For a Dirichlet boundary condition, we have a prescribed value on the boundary, $f_b$, which may be a function of time and/or space.
Given uniform spacing in $y$, we also have $l_2 + l_3 = l_1 = dy$.
The question then is given $l_{1,2,3}, f_{1,2,b}$, what is $f_2'$?

We should like to avoid adapting the finite difference scheme at each point which interacts with the boundary as above in order to keep the implementation as simple as possible.
One possible solution is to fill in the value of the field on the ``leg'' of the field line, $f_3$, and then use the standard finite difference scheme to compute the parallel derivatives.
We call this scheme ``Leg Value Fill'' (LVF).
This involves an extrapolation which needs to be accurate enough to not degrade the accuracy of the FD scheme.

We start from the Taylor expansions, truncated to third order, of $f_1, f_2, f_3$ about the boundary:
\begin{align}
  f_1 &= f_b - (l_1 + l_2)f_b' + \frac{1}{2}(l_1 + l_2)^2f_b'' - \frac{1}{6}(l_1 + l_2)^3f_b''', \label{eq:f1} \\
  f_2 &= f_b - l_2f_b' + \frac{1}{2}l_2^2f_b'' - \frac{1}{6}l_2^3f_b''', \label{eq:f2} \\
  f_3 &= f_b + l_3f_b' + \frac{1}{2}l_3^2f_b'' + \frac{1}{6}l_3^3f_b'''. \label{eq:f3}
\end{align}

We then use \cref{eq:f3,eq:f2} to get the first derivative at the boundary
\begin{align}
  f_b' &= \frac{1}{l_2l_3^2 + l_2^2l_3}[l_2^2f_3 + (l_3^2 - l_2^2)f_b - l_3^2f_2] - \frac{1}{6}l_2l_3f_b''' \label{eq:fb-prime}.
\end{align}
As $f_b'''$ is unknown, this is second-order accurate.
Similarly, we can also get the second derivative:
\begin{align}
  f_b'' = \frac{2[l_2f_3 - (l_2 + l_3)f_b + l_3f_2]}{l_3l_2^2 + l_2l_3^2} - \frac{(l_3^2 - l_2^2)}{3(l_2 + l_3)}f_b''' + ... \label{eq:fb-prime2}
\end{align}
The error in this expression is first order, except for the special case where $l_2 = l_3$ and \cref{eq:fb-prime2} reduces to the standard central difference scheme.

We can combine \cref{eq:f3,eq:f2}:
\begin{equation}
  f_b = \frac{l_2f_3 + l_3f_2}{l_2 + l_3} + \frac{f_b''}{2}\frac{l_2l_3^2 + l_2^2l_3}{l_2 + l_3} + ...
\end{equation}
Note that the error term ($f_b''$) is second order in the $l_{1,2,3}$ lengths, so the value at the boundary is determined to second order accuracy.
In order to do this, $f_3$ must be set to
\begin{equation}
  \label{eq:LVF}
  f_3 = f_b\frac{l_2 + l_3}{l_2} - \frac{l_3}{l_2}f_2 - \frac{f_b''}{2}\frac{l_2l_3^2 + l_2^2l_3}{l_2 + l_3} + ...
\end{equation}
This result can then be used in an arbitrary finite difference scheme to give the parallel derivatives of $f_2$.

For example, putting \cref{eq:LVF} into the standard \nth{2}-order accurate central difference for the first derivative:
\begin{equation}
  \label{eq:LVF-CD}
  f_2' = \frac{f_b\tfrac{l_1}{l_2} + f_2(1 - \tfrac{l_1}{l_2}) - f_1}{2l_1} - \frac{f_b''}{2}\frac{l_2l_3^2 + l_2^2l_3}{l_2 + l_3} + ...
\end{equation}
It can be seen that this result is still second-order in $l_1, l_2, l_3$.

We can actually go further and get a \nth{3}-order accurate scheme.
Insert \cref{eq:fb-prime,eq:fb-prime2} into \cref{eq:f1}:
\begin{align}
  f_1 &= -f_b\frac{l_1(l_1 + l_2 + l_3)(l_2 + l_3)}{l_2^2l_3 + l_2l_3^3} \nonumber \\
      &+ f_2 \frac{l_3}{l_2^2l_3 + l_2l_3^2}[(l_1 + l_2)l_3 + (l_1 + l_2)^2] \nonumber \\
      &+ f_3 \frac{(l_1^2l_2 + l_1l_2^2)}{l_2^2l_3 + l_2l_3^2} \nonumber \\
      &- f_b'''\frac{1}{6}[l_1^2(l_1 + l_2 + l_3) + 2l_1l_2l_3 + l_2^2l_3],   \label{eq:f1-2}
\end{align}
drop the $f_b'''$ term and rearrange for $f_3$:
\begin{align}
  f_3 &= \frac{l_2^2l_3 + l_2l_3^2}{(l_1^2l_2 + l_1l_2^2)}f_1 \nonumber \\
      &+ f_b\frac{l_1(l_1 + l_2 + l_3)(l_2 + l_3)}{(l_1^2l_2 + l_1l_2^2)} \nonumber \\
      &- f_2 \frac{l_3}{(l_1^2l_2 + l_1l_2^2)}[(l_1 + l_2)l_3 + (l_1 + l_2)^2].   \label{eq:f3-2}
\end{align}
$f_3$ is now known to third order, and can again be inserted into a standard finite difference scheme.

These two schemes, for second- and third-order, use the points along the field line which are already used in the second-order FCI parallel derivative operator.
Higher order schemes can be derived along similar lines, but these require more points along the field lines.
These could be generated at the same time as the initial field line tracing.

It is natural to ask if this scheme has consequences for field lines that intersect the boundary at shallow angles, or equivalently with low perpendicular resolution grids.
Magnetic field lines might be so shallow as to intersect many perpendicular planes before hitting the boundary.
That is, the intersection point on the adjacent plane may be outside the grid but still inside the material wall.
We don't anticipate this to be a problem, as for this case, the LVF scheme changes from an extrapolation in the parallel direction, to an interpolation which is often more numerically stable.
This can be seen by reducing the tilt of the field line in \cref{fig:problem-desc}.
When the field line is angled such that $f_3$ now lies above the boundary (but still below $f_2$ in the perpendicular direction), then $f_b$ must be now further along the field line from $f_2$ than $f_3$.

We have also derived an expression for $f_2'$ based on a non-uniform grid.
Instead of extrapolating to find values on the field line ``leg'', one can use the value on the boundary directly, but now the finite difference scheme for the parallel derivative must be adapted in order to maintain the second order accuracy.
The second order accurate central difference for parallel derivative using this scheme is
\begin{align}
  \label{eq:non-uniform-CD}
  f_2' = \frac{f_b\tfrac{l_1}{l_2} + f_2(\tfrac{l_2}{l_1} - \tfrac{l_1}{l_2}) - \tfrac{l_2}{l_1}f_1}{l_1 + l_2}.
\end{align}
However, when we tested this approach in a python toy model, we found that this scheme was more prone to numerical instabilities.

Further boundary condition schemes have also been investigated, such as asymmetric or one-sided differences.
For these types of schemes, the field line needs to be traced further to the two immediately adjacent poloidal slices.
However, the LVF scheme appears to demonstrate the best numerical properties and is the simplest to implement.

\subsubsection{Arbitrary geometry}
\label{sec:arb-geom}

The boundary scheme presented here is well-suited to a logical rectangular mesh, or the case where limiters are infinitesimally thin and so do not present a face to the magnetic field in the perpendicular direction, or mask the perpendicular grid.
While this scheme also works in the case of more complex material boundaries, the problem is a more general one of how to represent the material geometry numerically.
The mesh has to either follow the geometry, or grid cells must be ``masked'' where they intersect the material walls and the equations not evolved there.
A masked mesh complicates not just the interpolation for the LVF boundary scheme, but also perpendicular operators and boundary conditions.

Currently, \bout{} uses a logical rectangular mesh with optional branch cuts to handle X-points.
Recent work\cite{Leddy2016} has enabled this grid to follow the material boundaries more accurately.
Future work to upgrade \bout{} will also explore grids which can handle complex machine geometries, building on the work presented here.

\subsection{Implementation}
\label{sec:implementation}

The derivation of the FCI technique is discussed in Refs.~\onlinecite{Hariri2013,Hariri2014}; here we discuss its particular implementation in \bout{}.
There are three major steps required for FCI: first, the magnetic field lines must be followed from each grid point in both directions, and the intersection points with the adjacent perpendicular planes recorded; secondly, the scalar field must be interpolated at the intersection points; lastly, a finite difference can be applied using the interpolated values.

Following the magnetic field lines, or \emph{field line tracing}, generates a \emph{field line map} that maps a given grid point to its intersection point on the next/previous perpendicular plane.
Two field line maps are needed, one for the forwards (positive $y$) and one for the backwards (negative $y$) directions.
We construct these field line maps with a tool called \zoidberg{}, written in python.
\zoidberg{} uses \emph{odeint} from \emph{SciPy}\cite{Jones} to trace the field lines.
The magnetic field can be supplied to \zoidberg{} either as an tuple of three analytic functions (for $B_x(x, y, z), B_y(x, y, z), B_z(x, y, z)$), or a tuple of arrays which are to be interpolated by \emph{odeint}.
The latter form allows general numeric equilibria (from e.g. \emph{VMEC}\cite{Hirshman1983} or \emph{EFIT}\cite{Lao1985} files) to be used as input for FCI grids in \bout{}.
The output from \zoidberg{} is a file containing the field line maps.
This is an input to \bout{} -- currently, only time-independent magnetic fields are supported.

The second step of the FCI method, interpolation, is handled internally in \bout{}.
At each time-step, all fields which are to be acted upon by parallel derivative operators must be interpolated at the points held in the field line maps.
For details of the specific interpolation techniques used in \bout{}, see \cref{sec:interp}.

The boundary conditions in \bout{} are set at run-time, including the choice of making the $y$ and/or $z$ boundaries periodic for FCI.
Currently, non-periodic $z$ boundaries are only supported by the FCI parallel derivative operators in \bout{}, and not by any other spatial operator.
Future work will address supporting non-periodic $z$ boundaries generically.

During the initialisation stage in \bout{}, the field line maps are read in, the field lines that hit the edge are detected, and for each such field line a data structure of information required for the boundary condition is appended to a vector.
A separate vector of these structures is kept for the forward and backward directions, and each vector is stored in a \emph{BoundaryRegionPar} class.
When the boundary conditions are applied to a field during the course of the simulation, these vectors can be iterated over, and the LVF scheme is applied to populate the relevant points.

The \emph{BoundaryRegionPar} class needs to know some pieces of information about the field lines that intersect the boundaries.
These are the originating index point, the index-space coordinates of the intersection with the boundary, and the angle and distance to the boundary.
Briefly, the algorithm to collect this information is implemented as follows:
first determine which, if any, edges the field lines intersect; then find the coordinates of the intersection point.
For simple, planar boundaries, determining the intersection point is a trivial application of trigonometry; for more complex boundaries, determining where the field lines intersect the material walls may need to be done in the field line tracing procedure.
In either case, once the intersection point with the boundary is determined, the distance along the field line, and the angle the field line makes to the boundary can be computed.
While the angle of intersection is not used in the present work, it may be useful in more sophisticated boundary conditions, e.g. Loizu\cite{Loizu2012} boundary conditions for plasma pre-sheaths in the divertor region of tokamaks, where the boundary ion velocity is proportional to the sine of the angle of intersection.

\subsection{Verification}
\label{sec:lvf-verify}

An important part of testing a numerical model is verifying that it correctly implements the mathematical model.
Validating that the mathematical model correctly represents reality is a separate consideration.
Given that it is often the case that an analytical solution cannot be constructed for a mathematical model, it is necessary to use a different technique, such as the Method of Manufactured Solutions\cite{Oberkampf2010,Salari2000,Roache1998} (MMS).
With MMS, an arbitrary ``manufactured'' solution is imposed, and the mathematical model is applied to this solution.
This manufactured solution is in general not an exact solution, however, the ``remainder'' may be added to the numerical model as source terms such that the manufactured solution now \emph{is} an exact solution of the modified model.
The error is defined as the difference between the numerical solution and the manufactured solution.
Details on how the MMS framework is implemented in \bout{} can be found in Ref.~\onlinecite{Dudson2016}.

\bout{}, including the FCI method, has been successfully verified using MMS in periodic domains\cite{Dudson2016}.
In this work we use MMS to verify the \nth{2}- and \nth{3}-order LVF boundary condition scheme, as well as to verify different interpolation methods (\cref{sec:interp}).
The same physics model, computational domain and magnetic field as in Ref.~\onlinecite{Dudson2016} were used, which we briefly restate here.
Two coupled differential equations were evolved for a single time-step:
\begin{equation}
  \label{eq:bout-model}
  \begin{split}
    \frac{\partial f}{\partial t} &= \nabla_\parallel g + D(dy)^2\nabla_\parallel^2 f \\
    \frac{\partial g}{\partial t} &= \nabla_\parallel f + D(dy)^2\nabla_\parallel^2 g
  \end{split}
\end{equation}
where $D=10$ is an artificial diffusivity used purely for numerical stability.
A sheared slab with dimensions $L_x = 0.1$m, $L_y = 10$m, $L_z = 1$m in the radial, parallel and binormal directions, respectively, and magnetic field $(B_x, B_y, B_z) = (0, 1, 0.05 + (x - 0.05)/10)$ was used.
The manufactured solution used was
\begin{align}
  f = \sin(\bar{y} - \bar{z}) + \cos(t)\sin(\bar{y} - 2\bar{z}), \label{eq:man-sol-f}\\
  g = \cos(\bar{y} - \bar{z}) - \cos(t)\sin(\bar{y} - 2\bar{z}), \label{eq:man-sol-g}
\end{align}
where $\bar{y}, \bar{z}$ are normalised to be between $0$ and $2\pi$.
The diffusion terms in \cref{eq:bout-model} scale with $dy^2$ and so do not affect the convergence of the error on $\nabla_\parallel$.
As in Ref.~\onlinecite{Dudson2016}, we scale the grid in $y$ and $z$ simultaneously.

\begin{figure}[htb]
  \centering
  \includegraphics[width=0.5\textwidth]{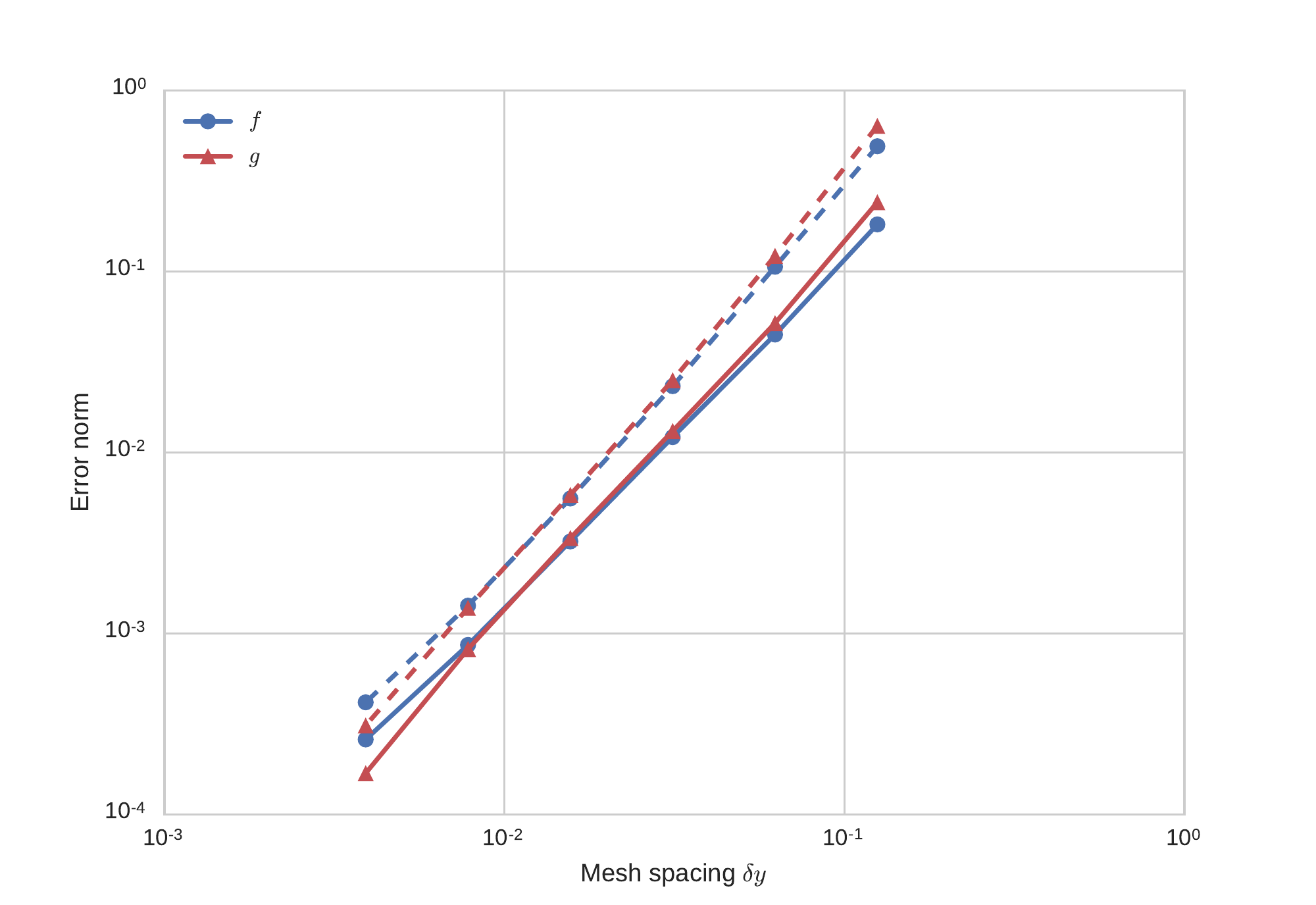}
  \caption{Error scaling for the \nth{2}-order LVF scheme in \bout{}. Solid lines are the $l_2$ norm, dashed lines are the $l_\infty$ norm (i.e. max. error).}
  \label{fig:bout-scaling-LVF2}
\end{figure}

\begin{figure}[htb]
  \centering
  \includegraphics[width=0.5\textwidth]{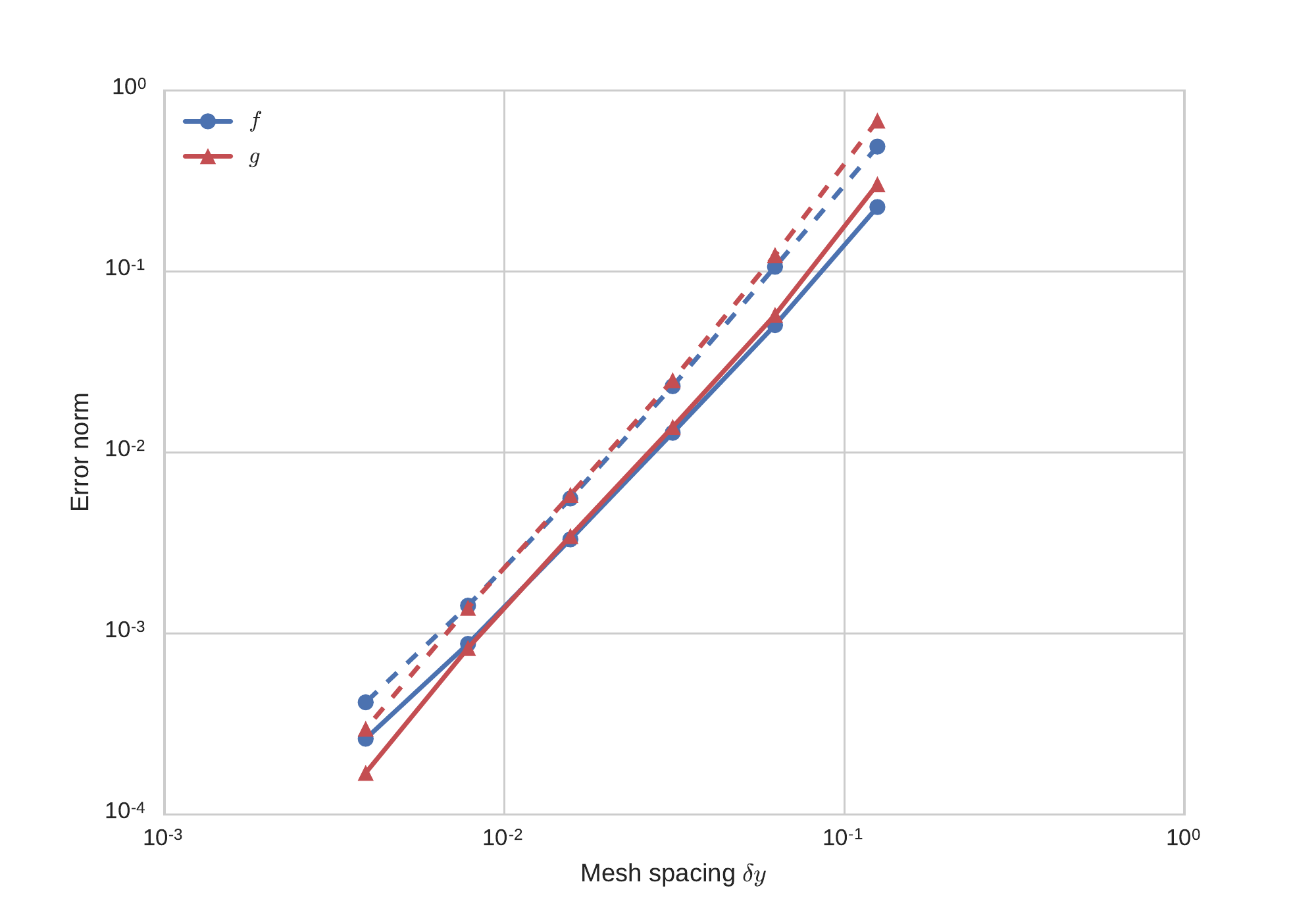}
  \caption{Error scaling for the \nth{3}-order LVF scheme in \bout{}. Solid lines are the $l_2$ norm, dashed lines are the $l_\infty$ norm (i.e. max. error).}
  \label{fig:bout-scaling-LVF3}
\end{figure}

\Cref{fig:bout-scaling-LVF2,fig:bout-scaling-LVF3} show the scaling of the MMS errors for $f, g$ for the \nth{2}- and \nth{3}-order LVF schemes, respectively, implemented in \bout{}.
The two schemes produce almost identical results, as the limiting factor on the error scaling is the finite difference scheme, which is second order.

It should be noted that because the \nth{3}-order LVF scheme relies on ``upstream'' information (i.e. points away from the boundary), it gets stuck in corners, where the field line leaves the boundary in both the forward and backward directions.
In these cases, the boundary condition cannot be applied, as is the case for the slab topology presented here.
As this is not possible, the results shown here are where the $z$-direction is periodic but the $y$-direction is not.
Switching which directions are periodic changes the order by only a fraction of a percent.


\subsection{Limiters}
\label{sec:limiters}

While true arbitrary shaped boundaries have not yet been implemented in \bout{} due to the reasons stated above, we have made the first steps by implementing an infinitesimally thin poloidal limiter.
Field lines either hit the limiter on the front/back face or they miss the limiter altogether and pass behind/in front of it.
Thus, no masking of the perpendicular grid is required, which would complicate operators in this plane.
The limiter is located halfway between the last and first $y$-planes.

Limiters are implemented in \bout{} as any function of $(x, z)$ (i.e. on the perpendicular plane) that passes through $0$, with positive values indicating the material walls.
This enables arbitrarily shaped limiters to be easily created.

The implementation of the limiter is very simple:
field lines that end on the $y=0$ slice can check if they hit the limiter by evaluating the limiter function described above.
If the result is positive, they are put in the same vectors of field lines held by the \emph{BoundaryRegionPar} objects, and treated identically.

\section{Interpolation}
\label{sec:interp}

The FCI method relies on interpolation in order to work, and it is the interpolation which is the most computationally expensive part of the technique (outside of the initial field line tracing, which only needs to be done once for static magnetic fields).
It is therefore important to understand how much of an impact the interpolation makes on the accuracy and efficency of the parallel derivate operator.
We have implemented three different interpolation methods - bilinear, four-point Lagrange and Hermite splines.
The choice of interpolation scheme is made at runtime.

After nearest-neighbour interpolation, bilinear interpolation is one of the most basic forms of interpolation in two dimensions, and consists of two sets of linear interpolation: first in one direction, then in the other.

Lagrange polynomials ensure that the interpolated function goes through the data points exactly.
Similarly to the bilinear interpolation, one dimensional polynomials are used to interpolate in each dimension successively.
An $n^\mathrm{th}$ order accurate scheme needs to use polynomials of degree at least $n$, which in turns requires at least $n+1$ data points.
Higher order polynomials can be used, but these are prone to over-fitting and spurious oscillations between the data-points.
A \nth{3}-order (4 point) 2D Lagrange interpolation is implemented in \bout{}.

Lastly, Hermite splines are piecewise polynomials that use the first derivative of the interpolant to act as a tension parameter, ensuring that the interpolated function is $C^1$ continuous.
Such splines are computationally more expensive than splines without tension parameters, as the first derivative needs to be evaluated several times for each interpolation.
A \nth{3}-order Hermite spline scheme is used in \bout{} as the default interpolation method for FCI.
This is the choice of interpolation scheme used in the original FCI papers\cite{Hariri2013,Hariri2014}.

\begin{figure}[ht]
  \centering
  \includegraphics[width=0.5\textwidth]{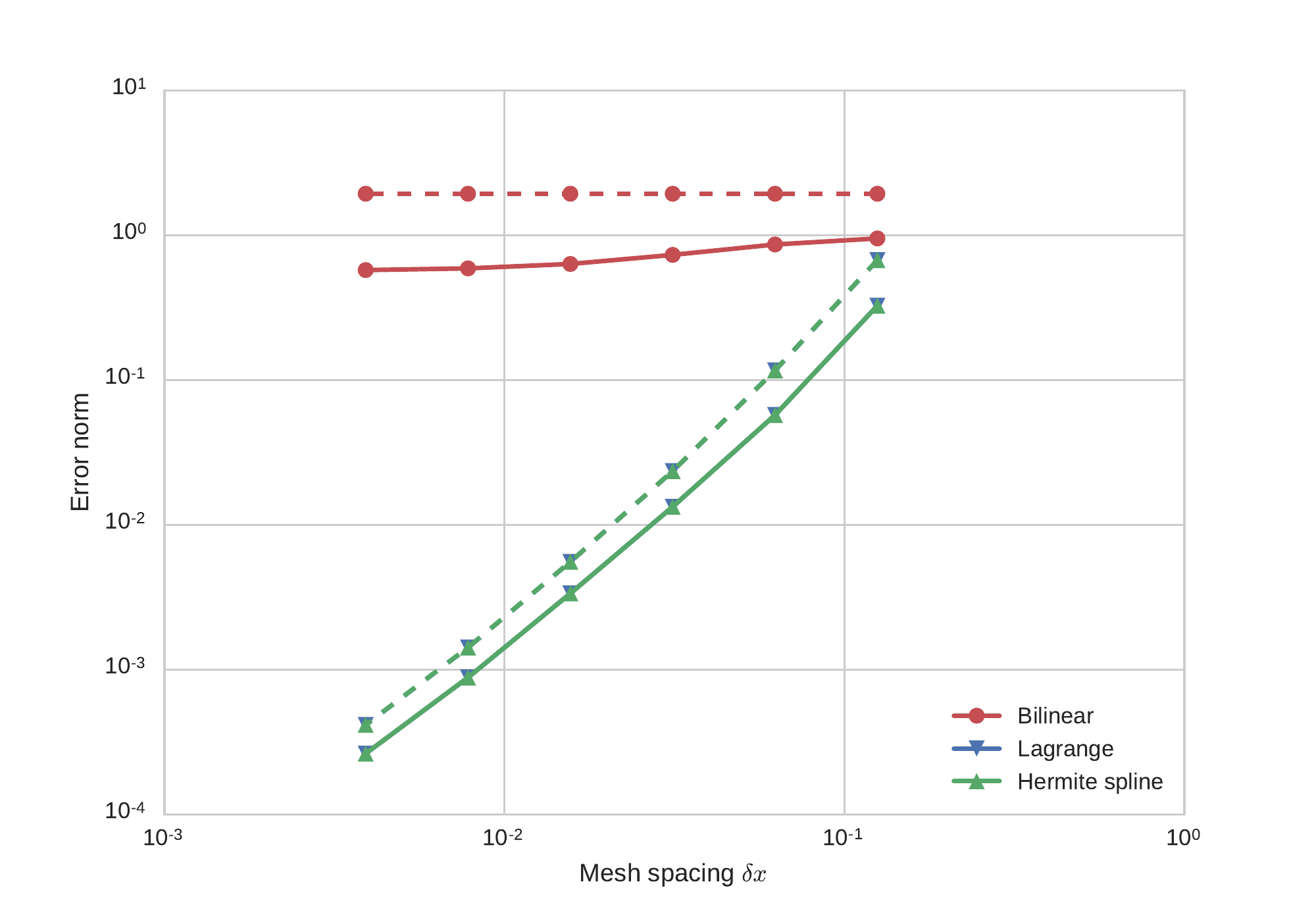}
  \caption{Error scaling of the field $f$ for three different interpolation schemes.
    Solid lines are the $l_2$ norm, dashed lines are the $l_\infty$ norm}
  \label{fig:interp-conv}
\end{figure}

We use the two-field wave model (\cref{eq:bout-model}), and verify the interpolation schemes using MMS (see \cref{sec:lvf-verify}).
The results are summarised in \cref{fig:interp-conv}.
Bilinear interpolation does not recover the expected scaling on $\nabla_\Vert$.
This is because the error on the interpolation is $O(dy)$, which is worse than the order of the finite difference scheme.
It is not clear why the overall scaling is then $O(1)$.

Four-point Lagrange and Hermite splines are both $O(dy^3)$, which is better than the finite difference error, and so recover the expected scaling.
The Hermite spline interpolation is roughly $\sim10\%$ more computationally expensive than the Lagrange polynomials due to the need to evaluate the first derivative.
However, it does ensure that the interpolated function is $C^1$ continuous, which may be advantageous, especially for non-linear simulations.

\section{3D magnetic fields}
\label{sec:3d-fields}

The FCI technique has been demonstrated and used in sheared slab\cite{Hariri2013}, cylindrical\cite{Hariri2013}, X-point and island\cite{Hill2015,Hariri2014}, and tokamak\cite{Stegmeir2016} magnetic geometries.
Here we demonstrate for the first time its use in stellarator-like fields.
This magnetic geometry is fully 3D, but has ``extrinsic'' curvature, i.e. the curvature has to be handled by a bracket operator in the physics model, rather than through the metric tensor.
Note that this is a limitation of the current version of \bout{}, and not of the FCI method in general.

\subsection{Stellarator geometry}
\label{sec:stellarator}

Due to the \bout{} limitations described above we implement a ``straight stellarator'', similar to a screw-pinch.
Because it's not possible to use a Grad-Shafranov solver for this magnetic equilibrium, we instead specify coils and compute $\myvec{B}$ from Amp\`ere's law.
We use four coils, defined by the position $\myvec{R}$ of the $k$-th coil which is:
\begin{align}
  \label{eq:coil-pos}
  \begin{split}
    \myvec{R_k}(\varphi) &= (x_0 + r_{coil}\cos(\tfrac{1}{2}k\pi + \iota\varphi))\hat{\myvec{x}} \\
    &+ (z_0 + r_{coil}\sin(\tfrac{1}{2}k\pi + \iota\varphi))\hat{\myvec{z}},
  \end{split}
\end{align}
where $(x_0,z_0)$ is the centre of the domain, $r_{coil}$ is the radius of the coil, $\iota$ is the rotational transform of the coils and the current in the $k$-th coil is given by
\begin{equation}
  \label{eq:coil-I}
  I_k = (-1)^{k} I_{coil},
\end{equation}
with $I_{coil}$ an input parameter.

The magnetic field at a point in space can then be computed as a sum of contributions from the coils:
\begin{align}
  \label{eq:coil-B}
  \begin{split}
    B_x(x,y,z) = \sum_{k=0} I_k \frac{C}{r_k^2}\sin(\theta_k) \\
    B_z(x,y,z) = \sum_{k=0} -I_k \frac{C}{r_k^2}\cos(\theta_k)
  \end{split}
\end{align}
where $r_k$ is the distance (in the $(x,z)$ plane) to the $k$-th coil, $\theta_k$ is the azimuthal angle to the coil, $C$ is some nature of constant.
We now have expressions for the magnetic field components which can be used as inputs to \zoidberg{} in order to trace the magnetic field and produce the field line maps required for \bout{}.

\begin{figure}[ht]
  \centering
  \includegraphics[width=0.5\textwidth]{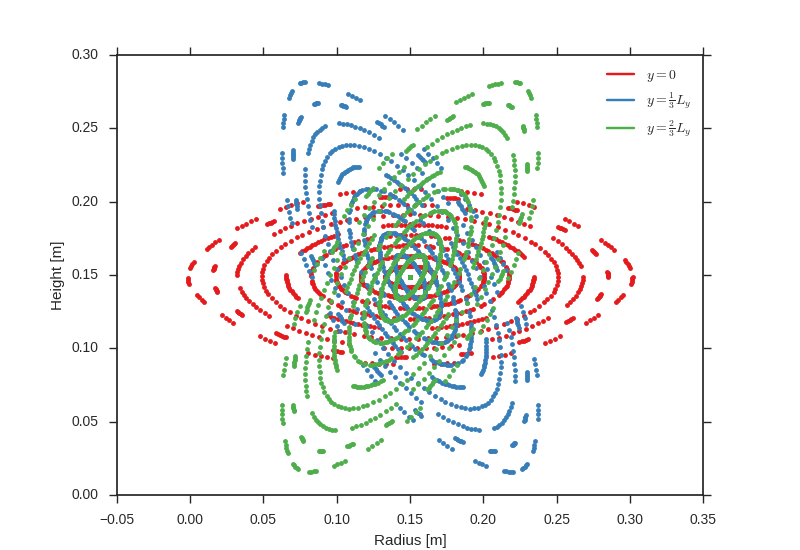}
  \caption{Poincar\'e plot of a straight stellarator at three $y$ planes}
  \label{fig:stellarator-poincare}
\end{figure}

\Cref{fig:stellarator-poincare} shows the Poincar\'e plot at three different $y$ locations, demonstrating the existence of flux surfaces.

We would also like to be able to initialise fields on flux surfaces.
While flux surfaces do exist for this magnetic topology, we do not have an expression for $\psi$, the poloidal magnetic flux.
Instead, we can use \zoidberg{} to construct numerical approximations to the flux surfaces.
By launching field lines from uniformly spaced radial positions, from the magnetic axis to one edge of the box, and by following them many times around the periodic domain in $y$, flux surfaces are eventually traced out.
Values in $[0, 1)$ are then assigned to the field lines according to their initial radial position, and these values then interpolated onto the simulation grid.
Points outside the last closed flux surface can be assigned the value $1$.
The resulting scalar field is numerical approximation to (normalised) $\psi$.
Initial conditions for the simulation fields can then be constructed in terms of this approximation to $\psi$ and are therefore flux functions, up to the accuracy of the field line tracing and the interpolation onto the grid.
The $\psi$ approximation is used only in the initialisation, and does not appear in the simulations.

\section{Simulations}
\label{sec:sims}

\subsection{Limiter}
\label{sec:limiter-sims}

We present here preliminary results showing how FCI is able to handle complex 3D magnetic geometry, including first steps towards arbitrary boundaries.
The magnetic geometry is a ``straight stellarator'' as described in \cref{sec:stellarator}.
The computational domain is a box, periodic in the $y$-direction, with Dirichlet boundary conditions in $(x, z)$.

For these initial simulations, we use a very simple parallel diffusion model:
\begin{equation}
  \label{eq:diffusion}
  \frac{\partial f}{\partial t} = D_\Vert\nabla_\parallel^2 f,
\end{equation}
where $f$ is some scalar field (which we refer to as density), and $D_\Vert$ is the parallel diffusivity.
Using this model, an initial perturbation will diffuse along the field lines, tracing out flux surfaces.
Due to the Dirichlet boundary conditions, density on field lines that hit the boundary will quickly decay away, effectively creating a last closed flux surface (LCFS).
Turning on the limiter will therefore change the position of the LCFS.

\begin{figure}[ht]
  \centering
  \includegraphics[width=0.5\textwidth]{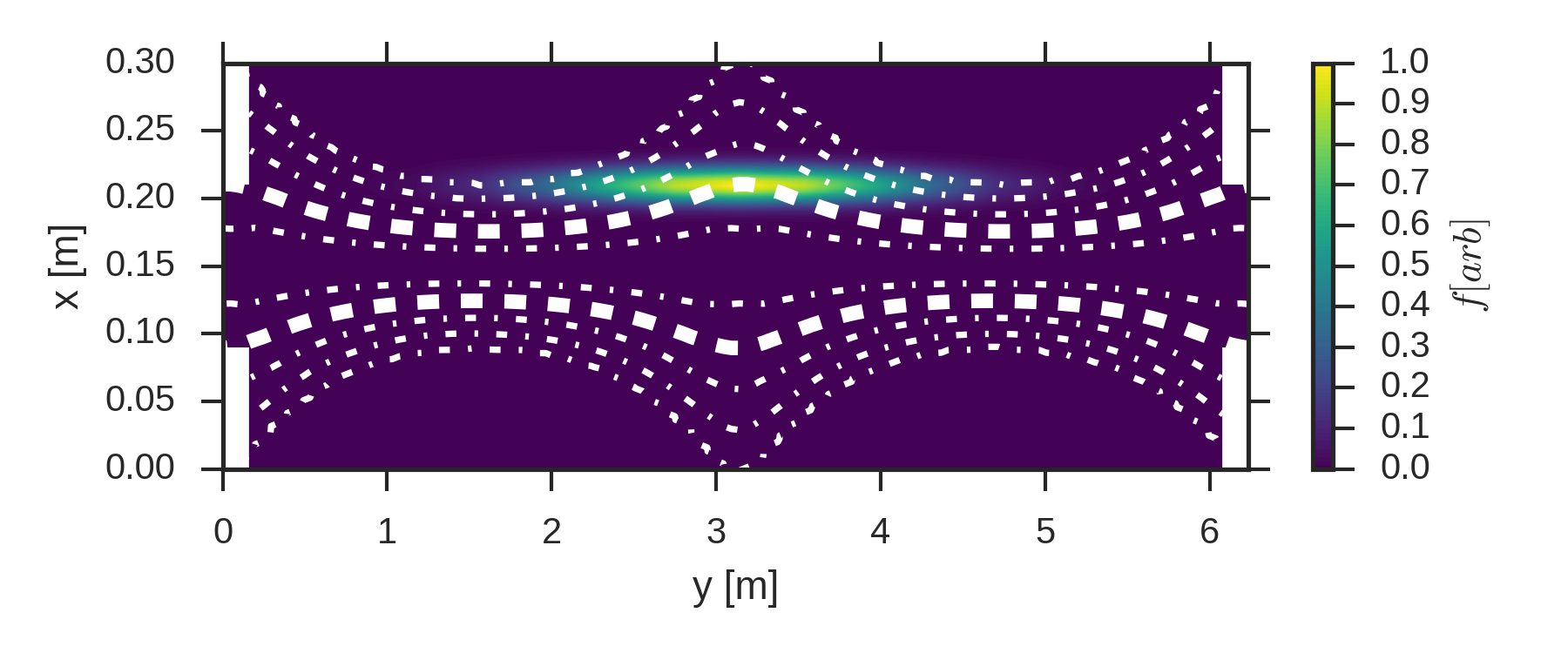}
  \caption{Heat map of initial condition for $f$ in diffusion model in the $(x, y)$ plane at $z=0.15$.
    Solid black lines indicate size and position of limiter.
    Dashed black lines indicate position of last closed flux surface.
    Dot-dashed lines show positions of flux surfaces.}
  \label{fig:diff-init}
\end{figure}

\Cref{fig:diff-init} shows the initial condition:
\begin{equation}
  \label{eq:diff-init}
  \begin{split}
    f(x, y, z; t=0) &= 100\gauss(x-0.2, 0.011) \\ 
    & \quad \times \gauss(z-0.15,0.3)\sin^6(\tfrac{1}{2}y),
  \end{split}
\end{equation}
where the normalised Gaussian with width $w$ is given by $\gauss(x, w) = \exp[-x^2 / (2w^2)] / (w\sqrt{2\pi})$.
The initial condition is a blob, spatially localised off-axis in $(x, z)$, with a wide distribution in $y$.

\begin{figure*}[ht]
  \centering
  \includegraphics[width=1\textwidth]{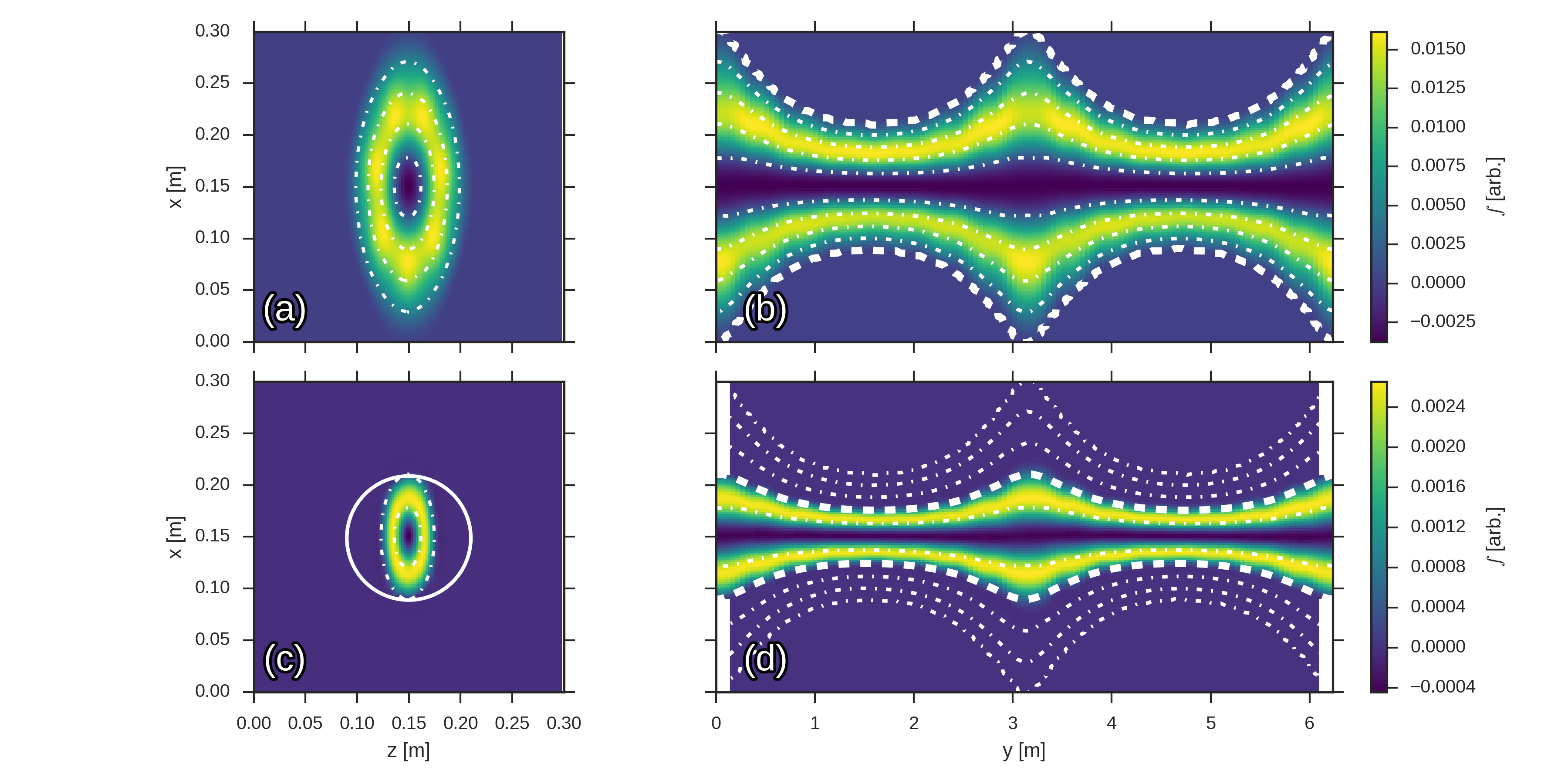}

    {\phantomsubcaption\label{fig:unlimit_x_vs_z}}
    {\phantomsubcaption\label{fig:unlimit_x_vs_y}}
    {\phantomsubcaption\label{fig:limit_x_vs_z}}
    {\phantomsubcaption\label{fig:limit_x_vs_y}}

  \caption{Heat map of $f$ in diffusion model at $t=400$ .
    (a): no limiter, (b): circular limiter.
    Solid black lines indicate size and position of limiter.
    Dashed black lines indicate position of last closed flux surface.
    Dot-dashed lines show positions of flux surfaces.
    Note that these figures have been upscaled in post-processing following the procedure outlined in \cref{sec:upscale}.}
  \label{fig:unlimit_limit_xy_xz}
\end{figure*}

The thin dot-dashed black lines in \cref{fig:diff-init} show the locations of flux surfaces.
In the absence of a limiter, the initial perturbation crosses most of the flux surfaces, whereas with a limiter, it is mostly outside the LCFS.
Snapshots of $f$ at late times, with and without a limiter, are shown in \cref{fig:unlimit_x_vs_y,fig:limit_x_vs_y}.
The density quickly diffuses along the field lines, either hitting the $(x, z)$ edges, or the limiter.
In either case, the field is cut off at the respective LCFS.

\Cref{fig:unlimit_limit_xy_xz} show the results of two simulations of the diffusion model (\cref{eq:diffusion}) at the same simulation time, \cref{fig:limit_x_vs_y,fig:limit_x_vs_z} have a circular limiter at $y=0$ centred on $x=0.15, z=0.15$ with radius $r=0.06$.
\Cref{fig:unlimit_x_vs_y,fig:limit_x_vs_y} show slices of the $(x,y)$ plane half-way through $z$, whereas \cref{fig:unlimit_x_vs_z,fig:limit_x_vs_z} are slices of the $(x,z)$ plane at $y=0$.
The vertical solid black lines in \cref{fig:limit_x_vs_y} and the solid black circle in \cref{fig:limit_x_vs_z} show the position of the limiter.
Note that the limiter is really infinitesimally thin, so presents surfaces only in the $(x,z)$ plane and has no $y$-extent.

\subsection{Upscaling}
\label{sec:upscale}

As with traditional field-aligned techniques, one of the \emph{raisons d'\^etre} of the FCI technique is the ability to use a low number of points in the parallel direction in order to resolve the relevant physics of a model.
Unfortunately, this has a downside when it comes to visualising the data.
Typically, visualisation programs use some nature of interpolation in the Cartesian (simulation grid) directions in order to show smoother images.
Because the magnetic field is not aligned with the grid, and structures in the data are typically aligned with the magnetic field, this results in rather blocky artefacts.
We can reduce or remove these artefacts by first upscaling, i.e. increasing the resolution in the parallel direction, the data ourselves.
If we assume the scalar field is slowing varying along the magnetic field line (which is an assumption of FCI itself), we can linearly interpolate along the field line to reconstruct the scalar field at higher parallel resolution.

The upscaling technique we use is as follows.
First, as with the usual FCI method, interpolate the data onto the field line end points in one direction.
Then, use a linear interpolation between the start and end points to get the desired number of additional points.
As well as interpolating the data, the $x, z$ displacements should also be linearly interpolated, which saves having to re-integrate the magnetic field.
We now have a ``cloud'' of data on new points.
Depending on the visualisation program, these new data can be interpolated themselves back onto a higher resolution rectangular grid, or left as a semi-unstructured grid.

\begin{figure}[ht]
  \centering
  \includegraphics[width=0.5\textwidth]{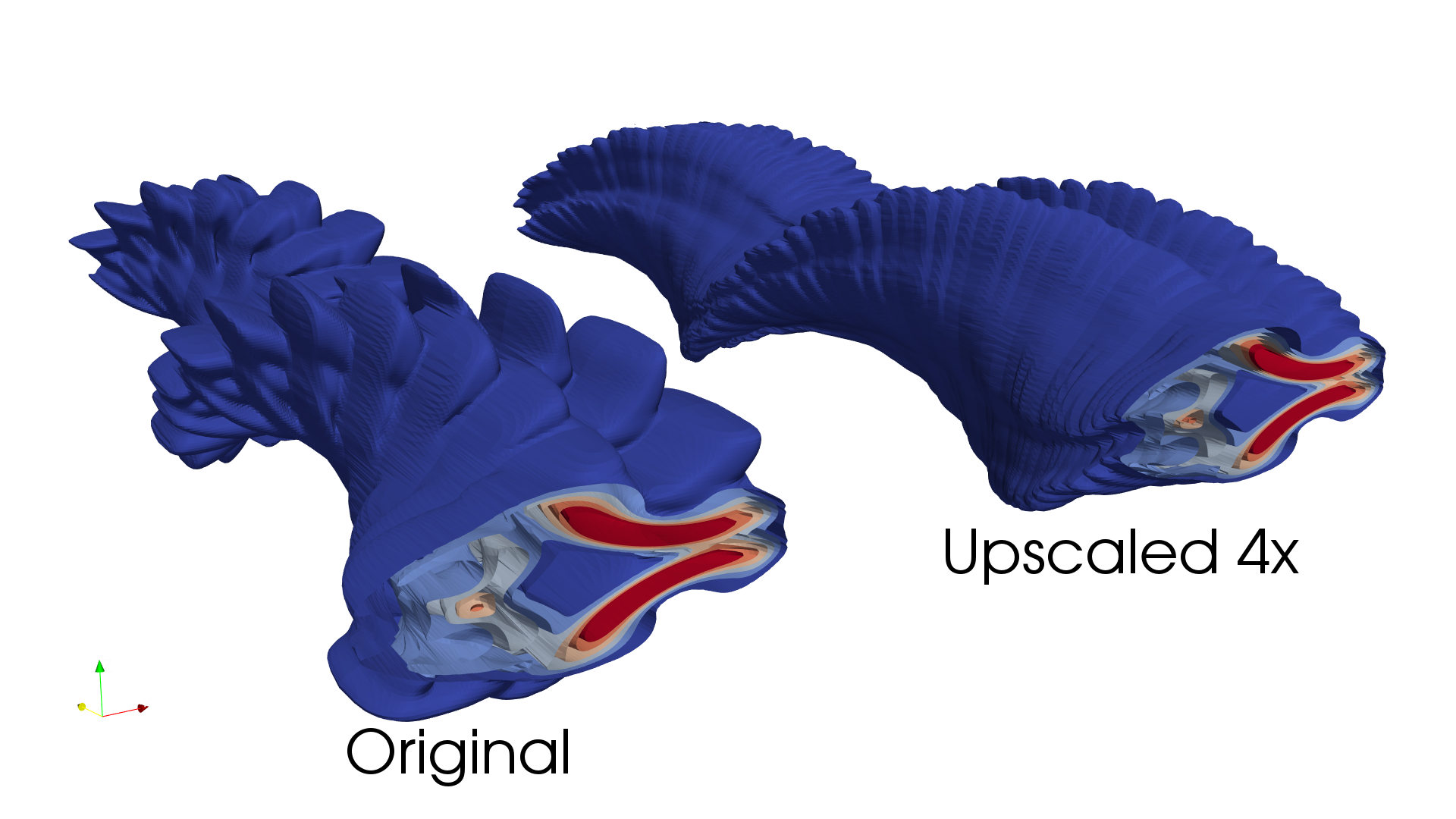}
  \caption{Visualisation of diffusion model (\cref{eq:diffusion}) in ParaView, original data on the left, upscaled by factor 4 on the right.
The presence of ``fins'' can be seen in the original (left) data.
These are caused by the visualisation program interpolating in the Cartesian directions, rather than along the magnetic field.
In the upscaled (right) version, the data has been interpolated in the parallel direction in order to reduce these fins.}
  \label{fig:upscale}
\end{figure}

\Cref{fig:upscale} contrasts the result of using this upscaling against the original data.
In the original data, there are clear unphysical lobes or fins which are aligned in the $y$ direction, although the simulation is well resolved.
In the upscaled version, there are still lobes, but they are now much smaller, and it is now easier to see how the density follows the field lines.

One issue with this upscaling algorithm is that it may give ``strange'' results when the data are not field-aligned, for example, as with initial conditions or injected sources.
In this case, the artefacts are now ``blocky'' in the parallel direction.
Note also that such structures will likely not be well resolved by either FCI or field-aligned approaches.

\subsection{Numerical diffusion}
\label{sec:num-diff}

There is some perpendicular (cross-field) diffusion from the numerical scheme, even in the model with only parallel derivatives (\cref{eq:diffusion}), due to, e.g. the interpolation scheme.
The numerical diffusion in FCI has already been characterised in axisymmetric magnetic geometries\cite{Hariri2013,Hariri2014,Stegmeir2016}.
Here we present an estimate of the numerical diffusion for the straight stellarator topology.
The expectation is that this should not be substantially different from the previous results\cite{Hariri2013}.

Using the diffusion model (\cref{eq:diffusion}) and initialising $f$ such that $\nabla_\Vert^2 f = 0$, the numerical diffusion can be estimated using
\begin{equation}
  \label{eq:num-diff}
  \frac{df}{dt} = D_\perp^{\textrm{eff}}\nabla_\perp^2 f
\end{equation}
where $D_\perp^{\textrm{eff}}$ is the effective perpendicular numerical diffusivity.
At each time-step, $df/dt$ and $\nabla_\perp^2 f$ can be saved and $D_\perp^\mathrm{eff}$ can be computed with
\begin{equation}
  \label{eq:num-diff-eff}
  D_\perp^{avg} = \langle \norm{\partial_tf} / \norm{\nabla_\perp^2f} \rangle,
\end{equation}
where $\norm{}$ is the 2-norm, and angle brackets indicate time average over latter half of simulation.
The time average is over the second half of the simulation in order to ignore the effect of initial transients.

In order to measure $D_\perp^{\textrm{eff}}$ we need to ensure that the parallel derivatives are zero, as this would appear to transport $f$ in the perpendicular plane.
To do this, $f$ must be initialised to a flux-function (i.e. constant on flux surfaces).
Because we do not have an expression for $\psi$, we must construct a numerical approximation to $\psi$ as described in \cref{sec:stellarator}, which can then be used to set an initial condition that is constant on flux surfaces.
The initial condition is a Gaussian in $\psi$,
\begin{equation}
  \label{eq:num-diff-init}
  f(\psi;t=0) = A\exp(-(\psi - \psi_0)^2 / (2\Delta^2)),
\end{equation}
with $A=1$, $\psi_0=0$ and $\Delta=0.1$.
Simulations were run up to $100t$, at fixed $n_y = 16$, with $n_x = n_z \in \{16, 32, 64, 128, 256\}$.

\begin{figure}[ht]
  \centering
  \includegraphics[width=0.5\textwidth]{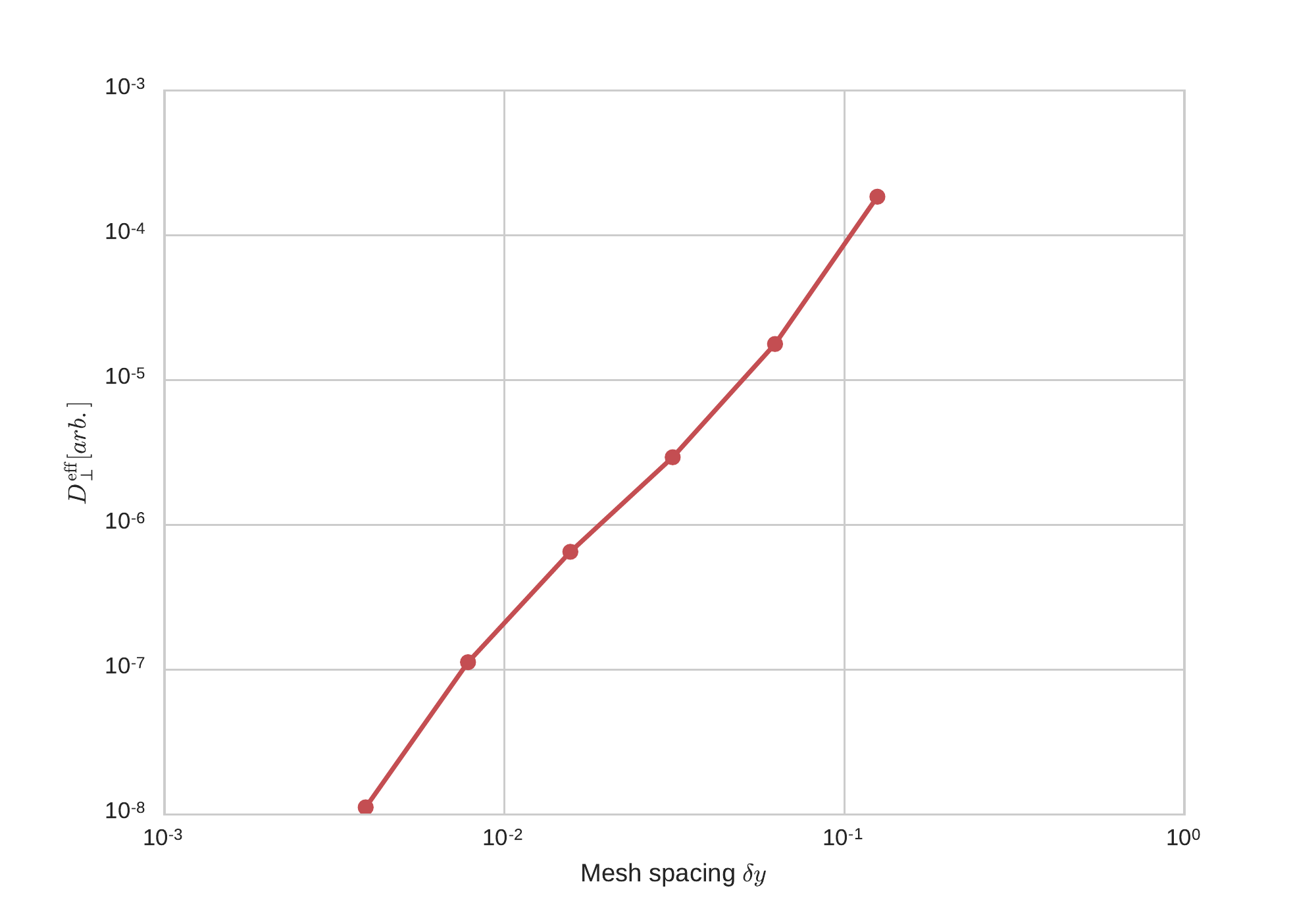}
  \caption{Numerical effective perpendicular diffusion as a function of mesh spacing.}
  \label{fig:num-diff}
\end{figure}

The results are summarised in \cref{fig:num-diff}.
The overall scaling of $D_\perp^{\textrm{eff}}$ with the perpendicular resolution is of order 2.67, and the absolute values are broadly in line with Ref.~\onlinecite{Hariri2013}, despite the magnetic topology there being axisymmetric.

\begin{figure*}[ht]
  \centering
  \begin{subfigure}[t]{0.5\linewidth}
    \centering
    \includegraphics[width=1.0\textwidth]{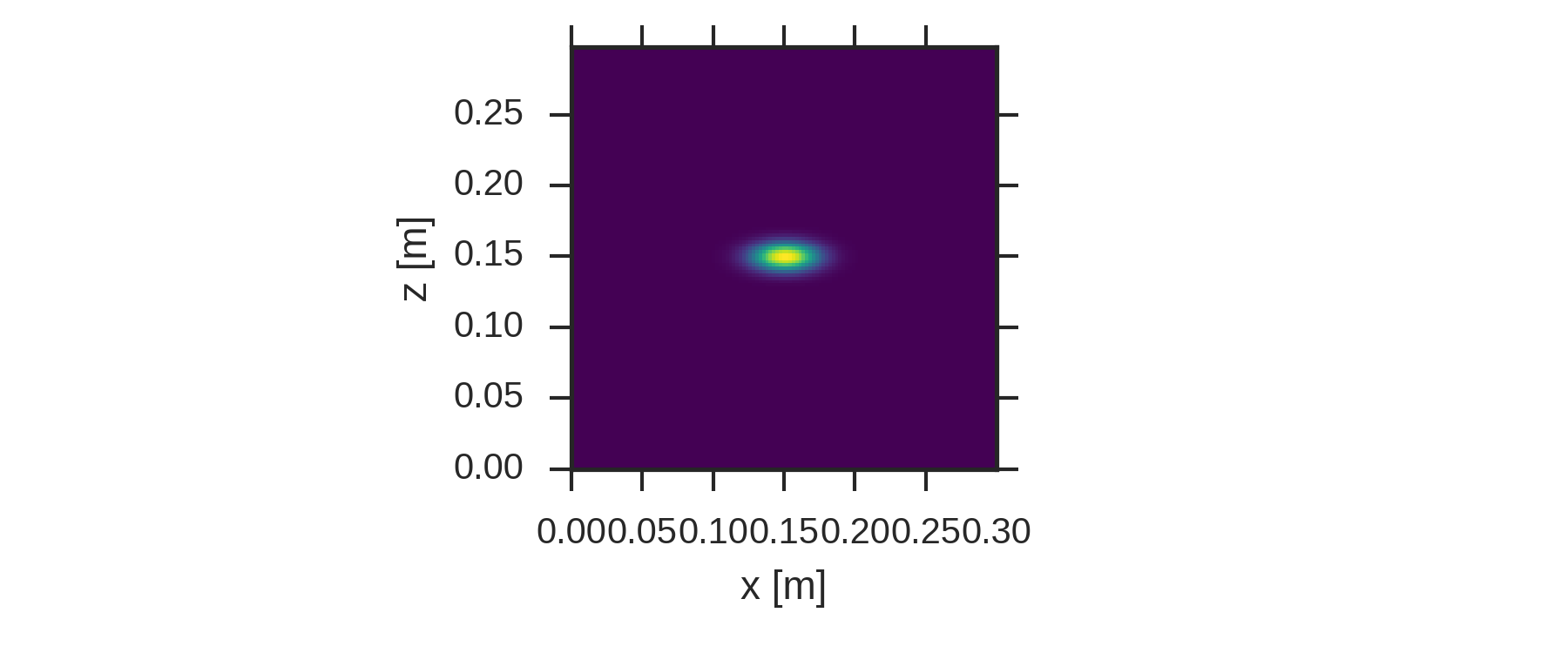}
    \caption{\label{fig:num-diff-init-xz}}
  \end{subfigure}%
  \begin{subfigure}[t]{0.5\linewidth}
    \centering
    \includegraphics[width=1.0\textwidth]{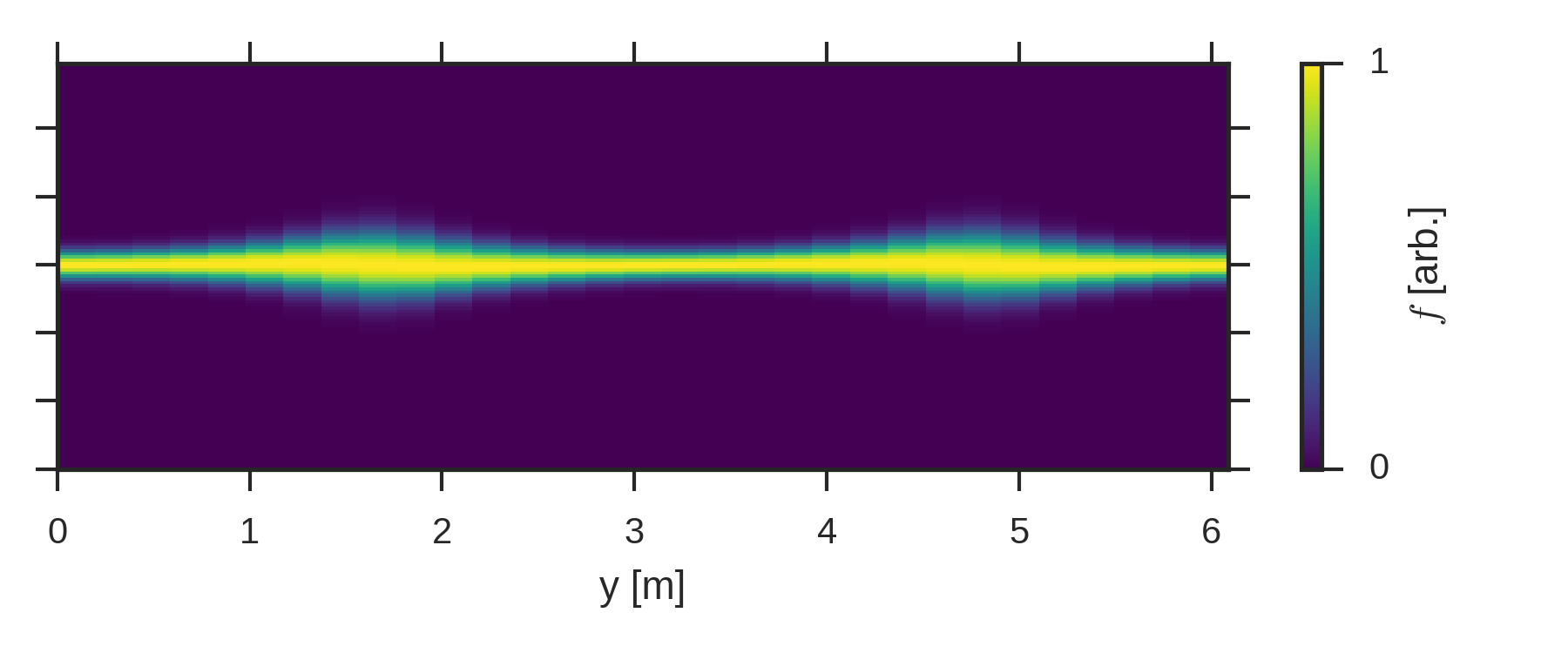}
    \caption{\label{fig:num-diff-init-yz}}
  \end{subfigure}
  \caption{Initial condition for the numerical diffusion test case. (a): $(x,z)$ plane at $y=0$, (b): $(y,z)$ plane at $x=0.15$.}

\end{figure*}

\section{Conclusions and discussion}
\label{sec:conclusion}

We have demonstrated a numerical scheme for parallel boundaries, where magnetic field lines intersect the material wall, for use with the Flux-Coordinate Independent (FCI) method for numerical derivatives parallel to the magnetic field.
The scheme for Dirichlet boundary conditions is based on a Taylor expansion about the boundary in order to extrapolate the field onto the ``leg'' of the field line outside the boundary.
Second- and third-order accurate versions of the scheme have been derived.
In the case of shallow grazing angles, where the field line intersects the next poloidal plane before the material wall, this scheme corresponds to an interpolation in the parallel direction, and so arbitrary-shaped material walls may be handled easily with the same scheme.
The Method of Manufactured Solutions (MMS) has been used to rigorously verify the accuracy and correct implementation of the boundary scheme.

The feasibility of performing simulations in non-axisymmetric magnetic configurations using the FCI method has been demonstrated, with a simple diffusion model in a straight, stellarator-like magnetic field.
An initial Gaussian blob in a simple diffusion model traces out flux surfaces.
The inclusion of a poloidal limiter reduces the radial extent of the flux surfaces thereby traced out.
Non-axisymmetry has been shown to not substantially affect the effective numerical diffusivity.

A novel technique for reducing blocky artefacts in visualisations during post-processing has also been demonstrated.
By linearly interpolating both the data to be visualised and the field line displacement map at the same time, the parallel resolution of the data can be up-sampled, and the new data re-interpolated onto a higher resolution grid.
Smoother, contours can then be produced, with fewer artefacts not present in the data.

An open question remains on the computational efficency of FCI.
Obviously, this does depend on the exact interpolation method used, the perpendicular grid resolution, the finite difference scheme, the degree of anisotropy in the physics, etc., but what is not obvious is when the cost of the FCI overheads is outweighed by the advantage in the parallel resolution.

An important consideration is that FCI is designed for complex magnetic topologies which are difficult to represent or capture with conventional field-aligned grids.
For example, the island divertors in a stellarator\cite{Grigull2001} involve multiple null points as well as large regions of stochastic magnetic field.
These would be very challenging to simulate using the usual mesh in \bout{}.
Another example would be the snowflake divertor concept\cite{Ryutov2007}, which has multiple legs.
This has been previously attempted in \bout{}\cite{Ma2014}, but this study was only able to capture the expanded flux surfaces in the region of the null point, and not the additional legs which are a feature of a second-order null point.
Here, then, it is clear that using the FCI method lets us get much further towards simulating plasma in these complex geometries, regardless of the computational cost.

In other situations, it is not so clear-cut that FCI presents a major advantage over a field-aligned grid.
Take, for instance, an island perturbation on a tokamak equilibrium.
This can be represented in a field-aligned grid simulation by splitting the magnetic field into equilibrium and perturbation caused by the island.
A bracket operator can then be used to capture the physics due to the island field.
The same method can be used for electromagnetic simulations where the perturbed magnetic field is a function of time.
In this case, further study is needed to determine the parameter regime where it is clearly advantageous to use FCI over a field-aligned grid. 

\subsection{Acknowledgements}
\label{sec:acknowledgements}

The authors would like to thank David Dickinson from the University of York for useful discussions and comments.

This work has been carried out with the framework of the EUROfusion Consortium and has received funding from the Euratom research and training programme 2014--2018 under grant agreement No. 633053.
The views and opinions expressed herein do not necessarily reflect those of the European Commission.

\bibliography{/home/peter/Documents/library.bib}

\begin{thebibliography}{28}%
\makeatletter
\providecommand \@ifxundefined [1]{%
 \@ifx{#1\undefined}
}%
\providecommand \@ifnum [1]{%
 \ifnum #1\expandafter \@firstoftwo
 \else \expandafter \@secondoftwo
 \fi
}%
\providecommand \@ifx [1]{%
 \ifx #1\expandafter \@firstoftwo
 \else \expandafter \@secondoftwo
 \fi
}%
\providecommand \natexlab [1]{#1}%
\providecommand \enquote  [1]{``#1''}%
\providecommand \bibnamefont  [1]{#1}%
\providecommand \bibfnamefont [1]{#1}%
\providecommand \citenamefont [1]{#1}%
\providecommand \href@noop [0]{\@secondoftwo}%
\providecommand \href [0]{\begingroup \@sanitize@url \@href}%
\providecommand \@href[1]{\@@startlink{#1}\@@href}%
\providecommand \@@href[1]{\endgroup#1\@@endlink}%
\providecommand \@sanitize@url [0]{\catcode `\\12\catcode `\$12\catcode
  `\&12\catcode `\#12\catcode `\^12\catcode `\_12\catcode `\%12\relax}%
\providecommand \@@startlink[1]{}%
\providecommand \@@endlink[0]{}%
\providecommand \url  [0]{\begingroup\@sanitize@url \@url }%
\providecommand \@url [1]{\endgroup\@href {#1}{\urlprefix }}%
\providecommand \urlprefix  [0]{URL }%
\providecommand \Eprint [0]{\href }%
\providecommand \doibase [0]{http://dx.doi.org/}%
\providecommand \selectlanguage [0]{\@gobble}%
\providecommand \bibinfo  [0]{\@secondoftwo}%
\providecommand \bibfield  [0]{\@secondoftwo}%
\providecommand \translation [1]{[#1]}%
\providecommand \BibitemOpen [0]{}%
\providecommand \bibitemStop [0]{}%
\providecommand \bibitemNoStop [0]{.\EOS\space}%
\providecommand \EOS [0]{\spacefactor3000\relax}%
\providecommand \BibitemShut  [1]{\csname bibitem#1\endcsname}%
\let\auto@bib@innerbib\@empty
\bibitem [{\citenamefont {Hariri}\ and\ \citenamefont
  {Ottaviani}(2013)}]{Hariri2013}%
  \BibitemOpen
  \bibfield  {author} {\bibinfo {author} {\bibfnamefont {F.}~\bibnamefont
  {Hariri}}\ and\ \bibinfo {author} {\bibfnamefont {M.}~\bibnamefont
  {Ottaviani}},\ }\href {\doibase 10.1016/j.cpc.2013.06.005} {\bibfield
  {journal} {\bibinfo  {journal} {Computer Physics Communications}\ }\textbf
  {\bibinfo {volume} {184}},\ \bibinfo {pages} {2419} (\bibinfo {year}
  {2013})}\BibitemShut {NoStop}%
\bibitem [{\citenamefont {Hariri}\ \emph {et~al.}(2014)\citenamefont {Hariri},
  \citenamefont {Hill}, \citenamefont {Ottaviani},\ and\ \citenamefont
  {Sarazin}}]{Hariri2014}%
  \BibitemOpen
  \bibfield  {author} {\bibinfo {author} {\bibfnamefont {F.}~\bibnamefont
  {Hariri}}, \bibinfo {author} {\bibfnamefont {P.}~\bibnamefont {Hill}},
  \bibinfo {author} {\bibfnamefont {M.}~\bibnamefont {Ottaviani}}, \ and\
  \bibinfo {author} {\bibfnamefont {Y.}~\bibnamefont {Sarazin}},\ }\href
  {\doibase 10.1063/1.4892405} {\bibfield  {journal} {\bibinfo  {journal}
  {Physics of Plasmas}\ }\textbf {\bibinfo {volume} {21}},\ \bibinfo {pages}
  {082509} (\bibinfo {year} {2014})}\BibitemShut {NoStop}%
\bibitem [{\citenamefont {Hill}, \citenamefont {Hariri},\ and\ \citenamefont
  {Ottaviani}(2015)}]{Hill2015}%
  \BibitemOpen
  \bibfield  {author} {\bibinfo {author} {\bibfnamefont {P.}~\bibnamefont
  {Hill}}, \bibinfo {author} {\bibfnamefont {F.}~\bibnamefont {Hariri}}, \ and\
  \bibinfo {author} {\bibfnamefont {M.}~\bibnamefont {Ottaviani}},\ }\href
  {\doibase 10.1063/1.4919031} {\bibfield  {journal} {\bibinfo  {journal}
  {Physics of Plasmas}\ }\textbf {\bibinfo {volume} {22}},\ \bibinfo {pages}
  {042308} (\bibinfo {year} {2015})}\BibitemShut {NoStop}%
\bibitem [{\citenamefont {Stegmeir}\ \emph {et~al.}(2016)\citenamefont
  {Stegmeir}, \citenamefont {Coster}, \citenamefont {Maj}, \citenamefont
  {Hallatschek},\ and\ \citenamefont {Lackner}}]{Stegmeir2016}%
  \BibitemOpen
  \bibfield  {author} {\bibinfo {author} {\bibfnamefont {A.}~\bibnamefont
  {Stegmeir}}, \bibinfo {author} {\bibfnamefont {D.}~\bibnamefont {Coster}},
  \bibinfo {author} {\bibfnamefont {O.}~\bibnamefont {Maj}}, \bibinfo {author}
  {\bibfnamefont {K.}~\bibnamefont {Hallatschek}}, \ and\ \bibinfo {author}
  {\bibfnamefont {K.}~\bibnamefont {Lackner}},\ }\href {\doibase
  10.1016/j.cpc.2015.09.016} {\bibfield  {journal} {\bibinfo  {journal}
  {Computer Physics Communications}\ }\textbf {\bibinfo {volume} {198}},\
  \bibinfo {pages} {139} (\bibinfo {year} {2016})}\BibitemShut {NoStop}%
\bibitem [{\citenamefont {Dudson}\ \emph {et~al.}(2009)\citenamefont {Dudson},
  \citenamefont {Umansky}, \citenamefont {Xu}, \citenamefont {Snyder},\ and\
  \citenamefont {Wilson}}]{Dudson2009}%
  \BibitemOpen
  \bibfield  {author} {\bibinfo {author} {\bibfnamefont {B.}~\bibnamefont
  {Dudson}}, \bibinfo {author} {\bibfnamefont {M.}~\bibnamefont {Umansky}},
  \bibinfo {author} {\bibfnamefont {X.}~\bibnamefont {Xu}}, \bibinfo {author}
  {\bibfnamefont {P.}~\bibnamefont {Snyder}}, \ and\ \bibinfo {author}
  {\bibfnamefont {H.}~\bibnamefont {Wilson}},\ }\href {\doibase
  10.1016/j.cpc.2009.03.008} {\bibfield  {journal} {\bibinfo  {journal}
  {Computer Physics Communications}\ }\textbf {\bibinfo {volume} {180}},\
  \bibinfo {pages} {1467} (\bibinfo {year} {2009})}\BibitemShut {NoStop}%
\bibitem [{\citenamefont {Dudson}\ \emph {et~al.}(2014)\citenamefont {Dudson},
  \citenamefont {Allen}, \citenamefont {Breyiannis}, \citenamefont {Brugger},
  \citenamefont {Buchanan}, \citenamefont {Easy}, \citenamefont {Farley},
  \citenamefont {Joseph}, \citenamefont {Kim}, \citenamefont {McGann},
  \citenamefont {Omotani}, \citenamefont {Umansky}, \citenamefont {Walkden},
  \citenamefont {Xia},\ and\ \citenamefont {Xu}}]{Dudson2014b}%
  \BibitemOpen
  \bibfield  {author} {\bibinfo {author} {\bibfnamefont {B.~D.}\ \bibnamefont
  {Dudson}}, \bibinfo {author} {\bibfnamefont {a.}~\bibnamefont {Allen}},
  \bibinfo {author} {\bibfnamefont {G.}~\bibnamefont {Breyiannis}}, \bibinfo
  {author} {\bibfnamefont {E.}~\bibnamefont {Brugger}}, \bibinfo {author}
  {\bibfnamefont {J.}~\bibnamefont {Buchanan}}, \bibinfo {author}
  {\bibfnamefont {L.}~\bibnamefont {Easy}}, \bibinfo {author} {\bibfnamefont
  {S.}~\bibnamefont {Farley}}, \bibinfo {author} {\bibfnamefont
  {I.}~\bibnamefont {Joseph}}, \bibinfo {author} {\bibfnamefont
  {M.}~\bibnamefont {Kim}}, \bibinfo {author} {\bibfnamefont {a.~D.}\
  \bibnamefont {McGann}}, \bibinfo {author} {\bibfnamefont {J.~T.}\
  \bibnamefont {Omotani}}, \bibinfo {author} {\bibfnamefont {M.~V.}\
  \bibnamefont {Umansky}}, \bibinfo {author} {\bibfnamefont {N.~R.}\
  \bibnamefont {Walkden}}, \bibinfo {author} {\bibfnamefont {T.}~\bibnamefont
  {Xia}}, \ and\ \bibinfo {author} {\bibfnamefont {X.~Q.}\ \bibnamefont {Xu}},\
  }\href {\doibase 10.1017/S0022377814000816} {\bibfield  {journal} {\bibinfo
  {journal} {Journal of Plasma Physics}\ }\textbf {\bibinfo {volume} {81}}
  (\bibinfo {year} {2014}),\ 10.1017/S0022377814000816},\ \Eprint
  {http://arxiv.org/abs/1405.7905} {arXiv:1405.7905} \BibitemShut {NoStop}%
\bibitem [{\citenamefont {Dudson}\ \emph {et~al.}(2016)\citenamefont {Dudson},
  \citenamefont {Madsen}, \citenamefont {Omotani}, \citenamefont {Hill},
  \citenamefont {Easy},\ and\ \citenamefont {L{\o}iten}}]{Dudson2016}%
  \BibitemOpen
  \bibfield  {author} {\bibinfo {author} {\bibfnamefont {B.}~\bibnamefont
  {Dudson}}, \bibinfo {author} {\bibfnamefont {J.}~\bibnamefont {Madsen}},
  \bibinfo {author} {\bibfnamefont {J.}~\bibnamefont {Omotani}}, \bibinfo
  {author} {\bibfnamefont {P.}~\bibnamefont {Hill}}, \bibinfo {author}
  {\bibfnamefont {L.}~\bibnamefont {Easy}}, \ and\ \bibinfo {author}
  {\bibfnamefont {M.}~\bibnamefont {L{\o}iten}},\ }\href {\doibase
  10.1063/1.4953429} {\bibfield  {journal} {\bibinfo  {journal} {Physics of
  Plasmas}\ }\textbf {\bibinfo {volume} {23}},\ \bibinfo {pages} {062303}
  (\bibinfo {year} {2016})},\ \Eprint {http://arxiv.org/abs/1602.06747}
  {arXiv:1602.06747} \BibitemShut {NoStop}%
\bibitem [{\citenamefont {Xi}\ \emph {et~al.}(2012)\citenamefont {Xi},
  \citenamefont {Xu}, \citenamefont {Wang},\ and\ \citenamefont
  {Xia}}]{Xi2012}%
  \BibitemOpen
  \bibfield  {author} {\bibinfo {author} {\bibfnamefont {P.~W.}\ \bibnamefont
  {Xi}}, \bibinfo {author} {\bibfnamefont {X.~Q.}\ \bibnamefont {Xu}}, \bibinfo
  {author} {\bibfnamefont {X.~G.}\ \bibnamefont {Wang}}, \ and\ \bibinfo
  {author} {\bibfnamefont {T.~Y.}\ \bibnamefont {Xia}},\ }\href {\doibase
  10.1063/1.4751256} {\bibfield  {journal} {\bibinfo  {journal} {Physics of
  Plasmas}\ }\textbf {\bibinfo {volume} {19}} (\bibinfo {year} {2012}),\
  10.1063/1.4751256}\BibitemShut {NoStop}%
\bibitem [{\citenamefont {Dudson}\ \emph {et~al.}(2011)\citenamefont {Dudson},
  \citenamefont {Xu}, \citenamefont {Umansky}, \citenamefont {Wilson},\ and\
  \citenamefont {Snyder}}]{Dudson2011}%
  \BibitemOpen
  \bibfield  {author} {\bibinfo {author} {\bibfnamefont {B.~D.}\ \bibnamefont
  {Dudson}}, \bibinfo {author} {\bibfnamefont {X.~Q.}\ \bibnamefont {Xu}},
  \bibinfo {author} {\bibfnamefont {M.~V.}\ \bibnamefont {Umansky}}, \bibinfo
  {author} {\bibfnamefont {H.~R.}\ \bibnamefont {Wilson}}, \ and\ \bibinfo
  {author} {\bibfnamefont {P.~B.}\ \bibnamefont {Snyder}},\ }\href {\doibase
  10.1088/0741-3335/53/5/054005} {\bibfield  {journal} {\bibinfo  {journal}
  {Plasma Physics and Controlled Fusion}\ }\textbf {\bibinfo {volume} {53}},\
  \bibinfo {pages} {054005} (\bibinfo {year} {2011})}\BibitemShut {NoStop}%
\bibitem [{\citenamefont {Snyder}\ \emph {et~al.}(2011)\citenamefont {Snyder},
  \citenamefont {Groebner}, \citenamefont {Hughes}, \citenamefont {Osborne},
  \citenamefont {Beurskens}, \citenamefont {a.W. Leonard}, \citenamefont
  {Wilson},\ and\ \citenamefont {Xu}}]{Snyder2011}%
  \BibitemOpen
  \bibfield  {author} {\bibinfo {author} {\bibfnamefont {P.}~\bibnamefont
  {Snyder}}, \bibinfo {author} {\bibfnamefont {R.}~\bibnamefont {Groebner}},
  \bibinfo {author} {\bibfnamefont {J.}~\bibnamefont {Hughes}}, \bibinfo
  {author} {\bibfnamefont {T.}~\bibnamefont {Osborne}}, \bibinfo {author}
  {\bibfnamefont {M.}~\bibnamefont {Beurskens}}, \bibinfo {author}
  {\bibnamefont {a.W. Leonard}}, \bibinfo {author} {\bibfnamefont
  {H.}~\bibnamefont {Wilson}}, \ and\ \bibinfo {author} {\bibfnamefont
  {X.}~\bibnamefont {Xu}},\ }\href {\doibase 10.1088/0029-5515/51/10/103016}
  {\bibfield  {journal} {\bibinfo  {journal} {Nuclear Fusion}\ }\textbf
  {\bibinfo {volume} {51}},\ \bibinfo {pages} {103016} (\bibinfo {year}
  {2011})}\BibitemShut {NoStop}%
\bibitem [{\citenamefont {Walkden}, \citenamefont {Dudson},\ and\ \citenamefont
  {Fishpool}(2013)}]{Walkden2013}%
  \BibitemOpen
  \bibfield  {author} {\bibinfo {author} {\bibfnamefont {N.~R.}\ \bibnamefont
  {Walkden}}, \bibinfo {author} {\bibfnamefont {B.~D.}\ \bibnamefont {Dudson}},
  \ and\ \bibinfo {author} {\bibfnamefont {G.}~\bibnamefont {Fishpool}},\
  }\href {\doibase 10.1088/0741-3335/55/10/105005} {\bibfield  {journal}
  {\bibinfo  {journal} {Plasma Physics and Controlled Fusion}\ }\textbf
  {\bibinfo {volume} {55}},\ \bibinfo {pages} {105005} (\bibinfo {year}
  {2013})},\ \Eprint {http://arxiv.org/abs/arXiv:1307.5234v1}
  {arXiv:arXiv:1307.5234v1} \BibitemShut {NoStop}%
\bibitem [{\citenamefont {Angus}, \citenamefont {Umansky},\ and\ \citenamefont
  {Krasheninnikov}(2012)}]{Angus2012}%
  \BibitemOpen
  \bibfield  {author} {\bibinfo {author} {\bibfnamefont {J.~R.}\ \bibnamefont
  {Angus}}, \bibinfo {author} {\bibfnamefont {M.~V.}\ \bibnamefont {Umansky}},
  \ and\ \bibinfo {author} {\bibfnamefont {S.~I.}\ \bibnamefont
  {Krasheninnikov}},\ }\href {\doibase 10.1103/PhysRevLett.108.215002}
  {\bibfield  {journal} {\bibinfo  {journal} {Physical Review Letters}\
  }\textbf {\bibinfo {volume} {108}},\ \bibinfo {pages} {1} (\bibinfo {year}
  {2012})}\BibitemShut {NoStop}%
\bibitem [{\citenamefont {Friedman}\ \emph {et~al.}(2012)\citenamefont
  {Friedman}, \citenamefont {Carter}, \citenamefont {Umansky}, \citenamefont
  {Schaffner},\ and\ \citenamefont {Dudson}}]{Friedman2012}%
  \BibitemOpen
  \bibfield  {author} {\bibinfo {author} {\bibfnamefont {B.}~\bibnamefont
  {Friedman}}, \bibinfo {author} {\bibfnamefont {T.~A.}\ \bibnamefont
  {Carter}}, \bibinfo {author} {\bibfnamefont {M.~V.}\ \bibnamefont {Umansky}},
  \bibinfo {author} {\bibfnamefont {D.}~\bibnamefont {Schaffner}}, \ and\
  \bibinfo {author} {\bibfnamefont {B.}~\bibnamefont {Dudson}},\ }\href
  {\doibase 10.1063/1.4759010} {\bibfield  {journal} {\bibinfo  {journal}
  {Physics of Plasmas}\ }\textbf {\bibinfo {volume} {19}} (\bibinfo {year}
  {2012}),\ 10.1063/1.4759010},\ \Eprint {http://arxiv.org/abs/1205.2337}
  {arXiv:1205.2337} \BibitemShut {NoStop}%
\bibitem [{\citenamefont {Shanahan}\ and\ \citenamefont
  {Dudson}(2014)}]{Shanahan2014}%
  \BibitemOpen
  \bibfield  {author} {\bibinfo {author} {\bibfnamefont {B.~W.}\ \bibnamefont
  {Shanahan}}\ and\ \bibinfo {author} {\bibfnamefont {B.~D.}\ \bibnamefont
  {Dudson}},\ }\href {\doibase 10.1088/1742-6596/561/1/012015} {\bibfield
  {journal} {\bibinfo  {journal} {Journal of Physics: Conference Series}\
  }\textbf {\bibinfo {volume} {561}},\ \bibinfo {pages} {012015} (\bibinfo
  {year} {2014})}\BibitemShut {NoStop}%
\bibitem [{\citenamefont {Xu}\ \emph {et~al.}(2008)\citenamefont {Xu},
  \citenamefont {Umansky}, \citenamefont {Dudson},\ and\ \citenamefont
  {Snyder}}]{Xu2008}%
  \BibitemOpen
  \bibfield  {author} {\bibinfo {author} {\bibfnamefont {X.~Q.}\ \bibnamefont
  {Xu}}, \bibinfo {author} {\bibfnamefont {M.~V.}\ \bibnamefont {Umansky}},
  \bibinfo {author} {\bibfnamefont {B.}~\bibnamefont {Dudson}}, \ and\ \bibinfo
  {author} {\bibfnamefont {P.~B.}\ \bibnamefont {Snyder}},\ }\href {\doibase
  10.4208/cicp.2008.v4.949} {\bibfield  {journal} {\bibinfo  {journal}
  {Communications in Computational Physics}\ }\textbf {\bibinfo {volume} {4}},\
  \bibinfo {pages} {949} (\bibinfo {year} {2008})}\BibitemShut {NoStop}%
\bibitem [{\citenamefont {Scott}(2001)}]{Scott2001}%
  \BibitemOpen
  \bibfield  {author} {\bibinfo {author} {\bibfnamefont {B.}~\bibnamefont
  {Scott}},\ }\href {\doibase 10.1063/1.1335832} {\bibfield  {journal}
  {\bibinfo  {journal} {Physics of Plasmas}\ }\textbf {\bibinfo {volume} {8}},\
  \bibinfo {pages} {447} (\bibinfo {year} {2001})}\BibitemShut {NoStop}%
\bibitem [{\citenamefont {Dimits}(1993)}]{Dimits1993}%
  \BibitemOpen
  \bibfield  {author} {\bibinfo {author} {\bibfnamefont {A.~M.}\ \bibnamefont
  {Dimits}},\ }\href {\doibase 10.1103/PhysRevE.48.4070} {\bibfield  {journal}
  {\bibinfo  {journal} {Physical Review E}\ }\textbf {\bibinfo {volume} {48}},\
  \bibinfo {pages} {4070} (\bibinfo {year} {1993})}\BibitemShut {NoStop}%
\bibitem [{\citenamefont {Leddy}\ \emph {et~al.}(2016)\citenamefont {Leddy},
  \citenamefont {Dudson}, \citenamefont {Romanelli}, \citenamefont {Shanahan},\
  and\ \citenamefont {Walkden}}]{Leddy2016}%
  \BibitemOpen
  \bibfield  {author} {\bibinfo {author} {\bibfnamefont {J.}~\bibnamefont
  {Leddy}}, \bibinfo {author} {\bibfnamefont {B.}~\bibnamefont {Dudson}},
  \bibinfo {author} {\bibfnamefont {M.}~\bibnamefont {Romanelli}}, \bibinfo
  {author} {\bibfnamefont {B.}~\bibnamefont {Shanahan}}, \ and\ \bibinfo
  {author} {\bibfnamefont {N.}~\bibnamefont {Walkden}},\ }\href@noop {} {\
  (\bibinfo {year} {2016})},\ \Eprint {http://arxiv.org/abs/arXiv:1604.05876v1}
  {arXiv:arXiv:1604.05876v1} \BibitemShut {NoStop}%
\bibitem [{\citenamefont {Jones}\ \emph {et~al.}(2001)\citenamefont {Jones},
  \citenamefont {Oliphant}, \citenamefont {Peterson},\ and\ \citenamefont
  {Others}}]{Jones}%
  \BibitemOpen
  \bibfield  {author} {\bibinfo {author} {\bibfnamefont {E.}~\bibnamefont
  {Jones}}, \bibinfo {author} {\bibfnamefont {T.}~\bibnamefont {Oliphant}},
  \bibinfo {author} {\bibfnamefont {P.}~\bibnamefont {Peterson}}, \ and\
  \bibinfo {author} {\bibnamefont {Others}},\ }\href {www.scipy.org} {\enquote
  {\bibinfo {title} {{SciPy: Open Source Scientific Tools for Python}},}\ }
  (\bibinfo {year} {2001})\BibitemShut {NoStop}%
\bibitem [{\citenamefont {Hirshman}\ and\ \citenamefont
  {Whitson}(1983)}]{Hirshman1983}%
  \BibitemOpen
  \bibfield  {author} {\bibinfo {author} {\bibfnamefont {S.~P.}\ \bibnamefont
  {Hirshman}}\ and\ \bibinfo {author} {\bibfnamefont {J.~C.}\ \bibnamefont
  {Whitson}},\ }\href {\doibase 10.1063/1.864116} {\bibfield  {journal}
  {\bibinfo  {journal} {Physics of Fluids}\ }\textbf {\bibinfo {volume} {26}},\
  \bibinfo {pages} {3553} (\bibinfo {year} {1983})}\BibitemShut {NoStop}%
\bibitem [{\citenamefont {Lao}\ \emph {et~al.}(1985)\citenamefont {Lao},
  \citenamefont {{St. John}}, \citenamefont {Stambaugh}, \citenamefont
  {Kellman},\ and\ \citenamefont {Pfeiffer}}]{Lao1985}%
  \BibitemOpen
  \bibfield  {author} {\bibinfo {author} {\bibfnamefont {L.}~\bibnamefont
  {Lao}}, \bibinfo {author} {\bibfnamefont {H.}~\bibnamefont {{St. John}}},
  \bibinfo {author} {\bibfnamefont {R.}~\bibnamefont {Stambaugh}}, \bibinfo
  {author} {\bibfnamefont {A.}~\bibnamefont {Kellman}}, \ and\ \bibinfo
  {author} {\bibfnamefont {W.}~\bibnamefont {Pfeiffer}},\ }\href {\doibase
  10.1088/0029-5515/25/11/007} {\bibfield  {journal} {\bibinfo  {journal}
  {Nuclear Fusion}\ }\textbf {\bibinfo {volume} {25}},\ \bibinfo {pages} {1611}
  (\bibinfo {year} {1985})}\BibitemShut {NoStop}%
\bibitem [{\citenamefont {Loizu}\ \emph {et~al.}(2012)\citenamefont {Loizu},
  \citenamefont {Ricci}, \citenamefont {Halpern},\ and\ \citenamefont
  {Jolliet}}]{Loizu2012}%
  \BibitemOpen
  \bibfield  {author} {\bibinfo {author} {\bibfnamefont {J.}~\bibnamefont
  {Loizu}}, \bibinfo {author} {\bibfnamefont {P.}~\bibnamefont {Ricci}},
  \bibinfo {author} {\bibfnamefont {F.~D.}\ \bibnamefont {Halpern}}, \ and\
  \bibinfo {author} {\bibfnamefont {S.}~\bibnamefont {Jolliet}},\ }\href
  {\doibase 10.1063/1.4771573} {\bibfield  {journal} {\bibinfo  {journal}
  {Physics of Plasmas}\ }\textbf {\bibinfo {volume} {19}} (\bibinfo {year}
  {2012}),\ 10.1063/1.4771573}\BibitemShut {NoStop}%
\bibitem [{\citenamefont {Oberkampf}\ and\ \citenamefont
  {Roy}(2010)}]{Oberkampf2010}%
  \BibitemOpen
  \bibfield  {author} {\bibinfo {author} {\bibfnamefont {W.~L.}\ \bibnamefont
  {Oberkampf}}\ and\ \bibinfo {author} {\bibfnamefont {C.~J.}\ \bibnamefont
  {Roy}},\ }\href@noop {} {\emph {\bibinfo {title} {{Verification and
  Validation in Scientific Computing}}}}\ (\bibinfo  {publisher} {Cambridge
  University Press},\ \bibinfo {address} {New York, NY},\ \bibinfo {year}
  {2010})\BibitemShut {NoStop}%
\bibitem [{\citenamefont {Salari}\ and\ \citenamefont
  {Knupp}(2000)}]{Salari2000}%
  \BibitemOpen
  \bibfield  {author} {\bibinfo {author} {\bibfnamefont {K.}~\bibnamefont
  {Salari}}\ and\ \bibinfo {author} {\bibfnamefont {P.}~\bibnamefont {Knupp}},\
  }\href@noop {} {\enquote {\bibinfo {title} {{Code Verification by the Method
  of Manufactured Solutions}},}\ }\bibinfo {type} {Tech. Rep.}\ (\bibinfo
  {institution} {Sandia National Laboratories},\ \bibinfo {year}
  {2000})\BibitemShut {NoStop}%
\bibitem [{\citenamefont {Roache}(1998)}]{Roache1998}%
  \BibitemOpen
  \bibfield  {author} {\bibinfo {author} {\bibfnamefont {P.~J.}\ \bibnamefont
  {Roache}},\ }\href@noop {} {\emph {\bibinfo {title} {{Verification and
  Validation in Computational Science and Engineering}}}}\ (\bibinfo
  {publisher} {Hermosa Publishers},\ \bibinfo {address} {Albuquerque NM},\
  \bibinfo {year} {1998})\BibitemShut {NoStop}%
\bibitem [{\citenamefont {Grigull}\ \emph {et~al.}(2001)\citenamefont
  {Grigull}, \citenamefont {McCormick}, \citenamefont {Baldzuhn}, \citenamefont
  {Burhenn}, \citenamefont {Brakel}, \citenamefont {Ehmler}, \citenamefont
  {Feng}, \citenamefont {Gadelmeier}, \citenamefont {Giannone}, \citenamefont
  {Hartmann}, \citenamefont {Hildebrandt}, \citenamefont {Hirsch},
  \citenamefont {Jaenicke}, \citenamefont {Kisslinger}, \citenamefont {Knauer},
  \citenamefont {K{\"{o}}nig}, \citenamefont {K{\"{u}}hner}, \citenamefont
  {Laqua}, \citenamefont {Naujoks}, \citenamefont {Niedermeye}, \citenamefont
  {Ramasubramanian}, \citenamefont {Rust}, \citenamefont {Sardei},
  \citenamefont {Wagner}, \citenamefont {Weller}, \citenamefont {Wenzel},\ and\
  \citenamefont {{W7-AS Team}}}]{Grigull2001}%
  \BibitemOpen
  \bibfield  {author} {\bibinfo {author} {\bibfnamefont {P.}~\bibnamefont
  {Grigull}}, \bibinfo {author} {\bibfnamefont {K.}~\bibnamefont {McCormick}},
  \bibinfo {author} {\bibfnamefont {J.}~\bibnamefont {Baldzuhn}}, \bibinfo
  {author} {\bibfnamefont {R.}~\bibnamefont {Burhenn}}, \bibinfo {author}
  {\bibfnamefont {R.}~\bibnamefont {Brakel}}, \bibinfo {author} {\bibfnamefont
  {H.}~\bibnamefont {Ehmler}}, \bibinfo {author} {\bibfnamefont
  {Y.}~\bibnamefont {Feng}}, \bibinfo {author} {\bibfnamefont {F.}~\bibnamefont
  {Gadelmeier}}, \bibinfo {author} {\bibfnamefont {L.}~\bibnamefont
  {Giannone}}, \bibinfo {author} {\bibfnamefont {D.}~\bibnamefont {Hartmann}},
  \bibinfo {author} {\bibfnamefont {D.}~\bibnamefont {Hildebrandt}}, \bibinfo
  {author} {\bibfnamefont {M.}~\bibnamefont {Hirsch}}, \bibinfo {author}
  {\bibfnamefont {R.}~\bibnamefont {Jaenicke}}, \bibinfo {author}
  {\bibfnamefont {J.}~\bibnamefont {Kisslinger}}, \bibinfo {author}
  {\bibfnamefont {J.}~\bibnamefont {Knauer}}, \bibinfo {author} {\bibfnamefont
  {R.}~\bibnamefont {K{\"{o}}nig}}, \bibinfo {author} {\bibfnamefont
  {G.}~\bibnamefont {K{\"{u}}hner}}, \bibinfo {author} {\bibfnamefont
  {H.}~\bibnamefont {Laqua}}, \bibinfo {author} {\bibfnamefont
  {D.}~\bibnamefont {Naujoks}}, \bibinfo {author} {\bibfnamefont
  {H.}~\bibnamefont {Niedermeye}}, \bibinfo {author} {\bibfnamefont
  {N.}~\bibnamefont {Ramasubramanian}}, \bibinfo {author} {\bibfnamefont
  {N.}~\bibnamefont {Rust}}, \bibinfo {author} {\bibfnamefont {F.}~\bibnamefont
  {Sardei}}, \bibinfo {author} {\bibfnamefont {F.}~\bibnamefont {Wagner}},
  \bibinfo {author} {\bibfnamefont {A.}~\bibnamefont {Weller}}, \bibinfo
  {author} {\bibfnamefont {U.}~\bibnamefont {Wenzel}}, \ and\ \bibinfo {author}
  {\bibnamefont {{W7-AS Team}}},\ }\href@noop {} {\bibfield  {journal}
  {\bibinfo  {journal} {Plasma Physics and Controlled Fusion}\ }\textbf
  {\bibinfo {volume} {43}},\ \bibinfo {pages} {A175} (\bibinfo {year}
  {2001})}\BibitemShut {NoStop}%
\bibitem [{\citenamefont {Ryutov}(2007)}]{Ryutov2007}%
  \BibitemOpen
  \bibfield  {author} {\bibinfo {author} {\bibfnamefont {D.~D.}\ \bibnamefont
  {Ryutov}},\ }\href {\doibase 10.1063/1.2738399} {\bibfield  {journal}
  {\bibinfo  {journal} {Physics of Plasmas}\ }\textbf {\bibinfo {volume} {14}}
  (\bibinfo {year} {2007}),\ 10.1063/1.2738399}\BibitemShut {NoStop}%
\bibitem [{\citenamefont {Ma}, \citenamefont {Xu},\ and\ \citenamefont
  {Dudson}(2014)}]{Ma2014}%
  \BibitemOpen
  \bibfield  {author} {\bibinfo {author} {\bibfnamefont {J.}~\bibnamefont
  {Ma}}, \bibinfo {author} {\bibfnamefont {X.}~\bibnamefont {Xu}}, \ and\
  \bibinfo {author} {\bibfnamefont {B.}~\bibnamefont {Dudson}},\ }\href
  {\doibase 10.1088/0029-5515/54/3/033011} {\bibfield  {journal} {\bibinfo
  {journal} {Nuclear Fusion}\ }\textbf {\bibinfo {volume} {54}},\ \bibinfo
  {pages} {033011} (\bibinfo {year} {2014})}\BibitemShut {NoStop}%
\end{thebibliography}%
\end{document}